\documentclass[twocolumn]{aastex631}

\let\splitbox\undefined
\usepackage{adjustbox}
\usepackage{rotating}
\usepackage{amsmath}

\usepackage{tablefootnote}

\begin{document}




\title{

The JWST Early Release Science Program for Direct Observations of Exoplanetary Systems VIII: patchy forsterite and enstatite clouds in the atmosphere of VHS 1256 b, retrieval lessons learned and outlook to the future}

\author[0000-0001-8818-1544]{Niall Whiteford}\thanks{Send manuscript correspondence to niallwhiteford@gmail.com}
\affiliation{Department of Astrophysics, American Museum of Natural History, Central Park West at 79th Street, New York, NY 10034, USA}

\author[0000-0001-6251-0573]{Jacqueline K. Faherty}
\affiliation{Department of Astrophysics, American Museum of Natural History, Central Park West at 79th Street, New York, NY 10034, USA}

\author[0000-0001-6251-0573]{Ben Burningham}
\affiliation{Centre for Astrophysics Research, Department of Physics, Astronomy and Mathematics, University of Hertfordshire, Hatfield AL10 9AB}

\author[0000-0003-0489-1528]{Johanna M. Vos}
\affiliation{School of Physics, Trinity College Dublin, The University of Dublin, Dublin 2, Ireland}
\affiliation{Department of Astrophysics, American Museum of Natural History, Central Park West at 79th Street, New York, NY 10034, USA}

\author[0000-0003-0331-3654]{Simon Petrus}\
\affiliation{N\'{u}cleo Milenio Formac\'{i}on Planetaria - NPF, Universidad de Valpara\'{i}so, Av. Gran Breta\~{n}a 1111, Valpara\'{i}so, Chile}
\affiliation{Millennium Nucleus on Young Exoplanets and their Moons (YEMS), Chile}
\affiliation{Departamento de F\'{i}sica, Universidad de Santiago de Chile, Av. Victor Jara 3659, Santiago, Chile}

\author[0000-0001-8718-3732]{Polychronis Patapis}
\affiliation{Institute of Particle Physics and Astrophysics, ETH Zurich, Wolfgang-Pauli-Str. 27, 8093 Zurich, Switzerland}

\author[0000-0003-4614-7035]{Beth A. Biller}
\affiliation{Scottish Universities Physics Alliance, Institute for Astronomy, University of Edinburgh, Blackford Hill, Edinburgh EH9 3HJ, UK}
\affiliation{Centre for Exoplanet Science, University of Edinburgh, Edinburgh EH9 3HJ, UK}

\author[0000-0001-6098-3924]{Andrew Skemer} 
\affiliation{University of California Santa Cruz, Santa Cruz, CA, USA}

\author[0000-0001-8074-2562]{Sasha Hinkley}
\affiliation{University of Exeter, Astrophysics Group, Physics Building, Stocker Road, Exeter, EX4 4QL, UK}

\author[0000-0001-8818-1544]{Emily Calamari}
\affiliation{Department of Astrophysics, American Museum of Natural History, Central Park West at 79th Street, New York, NY 10034, USA}

\author[0000-0002-2011-4924]{Genaro Suárez}
\affiliation{Department of Astrophysics, American Museum of Natural History, Central Park West at 79th Street, New York, NY 10034, USA}

\author[0000-0002-2011-4924]{Kelle L. Cruz}
\affiliation{Department of Physics and Astronomy, Hunter College, City University of New York, 695 Park Avenue, New York, NY 10065, USA}
\affiliation{Physics, Graduate Center of the City University of New York, 365 5th Avenue, New York, NY, 10016 USA}
\affiliation{Department of Astrophysics, American Museum of Natural History, Central Park West at 79th Street, New York, NY 10034, USA}

\author[0000-0002-5500-4602]{Brittany E. Miles}
\affiliation{University of California Santa Cruz, Santa Cruz, CA, USA}

\author[0000-0001-5365-4815]{Aarynn L. Carter}
\affiliation{University of California Santa Cruz, Santa Cruz, CA, USA}

\author[0000-0003-0150-3489]{Francisco A. Martinez}
\affiliation{Department of Astrophysics, American Museum of Natural History, Central Park West at 79th Street, New York, NY 10034, USA}

% ...




\begin{abstract}
JWST defines a new era for the data-driven approach of retrieval modelling, which has become a cornerstone tool for the statistical inference of exoplanetary and brown dwarf properties. The Early Release Science program \#1386 observations of VHS 1256 b represent a huge jump in data quality, data quantity and spectral coverage for such objects.  VHS \,1256 \,b is a young, planetary mass and extremely variable companion that populates the enigmatic L/T cohort of substellar atmospheres.  In this first retrieval analysis of the full 1 -- 18 $\mu$m dataset, we apply the $Brewster$ retrieval framework to the NIRSpec and MIRI spectroscopic observations of VHS 1256 b, exploring a variety of cloud species and structures. Using $\Delta$BIC we find that the data is best described by a forsterite (Mg$_2$SiO$_4$) and enstatite (MgSiO$_3$) cloud combination. Our analysis shows a strong preference for patchy silicate cloud coverage, which aligns with VHS 1256 b's extensive and well documented spectral variability. Our retrieval is able to place constraints on the abundances of H$_2$O, CO, CO$_2$, CH$_4$ as well as NH$_3$.
We also show that the retrieved parameters are sensitive to the data used and the relative signal-to-noise ratios between data from different instruments. 
We conclude with the next steps for the wider retrieval community to better understand young and cloudy exoplanetary atmospheres.  
\end{abstract}




\section{Introduction} \label{sec:intro}

Since the earliest detections of directly imaged exoplanets \citep{2004A&A_Chauvin, 2008Sci_Kalas, Marois2008} it became clear that they were spectral and characteristic siblings to brown dwarfs \citep{1995Natur_Rebolo, 1995_Nakajima, 1995_Oppenheimer}. Whilst separated by mass, the atmospheres of directly imaged exoplanets and brown dwarfs present the same or similar spectral signatures across a wide temperature range. This shows these two populations share a common evolutionary path in relation to the dominant chemistry, clouds and atmospheric processes present.   

This characteristic overlap led to the same atmospheric theory and modeling recipes which were first developed for brown dwarfs \citep[e.g.][]{1997_Burrows, 1999_Burrows_Sharp, 2001_Allard, AckermanMarley2001, 2002_Lodders_Fegley, 2006_Lodders_Chemistry_of_Low_Mass_Substellar_Objects, Saumon2006} then being applied to datasets for directly imaged exoplanets (or low surface gravity companions). This came in the form of evolutionary track or spectral grid model comparison \citep[e.g.][]{2008_Saumon_Marley_Evolution_models,2008_Fortney, Barman_2011_HR8799b, Barman_2011_2M1207b, 2011_Burningham_ross458c, Marley_2012_HR8799, Skemer_2012_hr8799, 2013_Konopacky, Liu_2013_PSOJ318, Bonnefoy_2014_betapicb, Macintosh2015, Chilcote_2015_betapicb}. These {self consistent grid frameworks, however, often lacked the model flexibility and freedom to successfully fit observations. The discrepancies between the models and data can be attributed to assumptions implemented in regards to the presence of chemical equilibrium vs disequilibrium, the Carbon/Oxygen (C/O) ratio and metallicity [M/H], the atmosphere's temperature-pressure structure and if -- or how -- clouds are implemented in the model. These are key atmospheric properties which still remain enigmatic. Without a constrained and robust understanding of these properties we cannot fully understand these objects, nor retrace how they formed, which is a key aim of exoplanet studies in the pursuit of contextualising our own solar system.

As such, retrieval analyses of exoplanet and brown dwarf spectra have become common practice in the past decade \citep[e.g.][]{Lee_2013_HR8799b_retrieval, Line_2015, Line_2017, Burningham2017, Lavie_2017_HELIOS, Zalesky_2019, Kitzmann_2020, Molliere_2020, Wang_2020, 2020_Gonzales,2021ApJ_Gonzales, Burningham_2021, 2022ApJ_Calamari, 2022ApJ_Lueber, 2022ApJ_Howe_apollo, 2023ApJ_Vos, 2023MNRAS_Whiteford, 2023Natur_Barrado, 2023AJ_Wang, 2024_Nasedkin_8799, 2024Natur_Faherty}. This has been accompanied by an explosion of retrieval tools available -- see \citet{2023RNAAS_MacDonald_list} for an extensive summary list of retrieval tools and their capabilities. This data-driven approach, via very flexible frameworks (which can vary tool-to-tool) can allow for improved modelling fits to observations of both these populations.

The challenge, however, has been doing deep-dive characterizations of the atmospheric and bulk properties of these objects as the essential combination of broadband spectral coverage (near and mid infrared) at high signal-to-noise (SNR, $\geq$100) has been limited.
This has been particularly true for the directly imaged exoplanets and companions which are plagued by observational challenges due to the presence of a much brighter host. For this reason these directly imaged exoplanets remained largely unmapped beyond 3$\mu$m apart from sparse low SNR spectra and photometric points \citep[e.g.][]{Miles_2018_PSOJ318, 2022AJ....163..217D_Stone}.

Another challenge with retrieval studies has been the inability to clearly and confidently present a single model conclusion, especially in regards to temperature-pressure structure and the presence of clouds \citep[e.g.][]{Molliere_2020,2022ApJ_Gonzales, 2023MNRAS_Whiteford}. Recently, \citet{Burningham_2021} and \citet{2023ApJ_Vos}  demonstrated the power of combining both near and mid-IR observations together to break this cloud vs temperature-pressure model degeneracy. This has permitted constraints to be placed on both cloud structures and condensate species.\citet{Burningham_2021} and \citet{2023ApJ_Vos} accomplished this by combining high-dimensional models in a retrieval framework (approximately 40 parameters) with spectra covering 1 - 14 $\mu$m of brown dwarfs in both the high and low surface gravity regimes. Wider wavelength coverage helps to inform the delicate, correlated and degenerate dance between temperature-pressure structure, the need to invoke cloud opacity, cloud particle size distributions and where to position these clouds. These correlated and degenerate relationship have been well documented for brown dwarfs and directly imaged exoplanets when wavelength coverage is limited to 1 -- 2.5 $\mu$m \citep[e.g.][]{Molliere_2020, 2023MNRAS_Whiteford, 2024arXiv_Phillips}.  Much wider wavelength coverage facilitates a much more complete coverage of the object's full spectral energy distribution (SED), access to more molecular features, and the signature of silicate clouds around 10 $\mu$m for the L dwarf regime \citep[e.g.][]{2006ApJ_Cushing, 2022MNRAS_Suarez}.

The James Webb Space Telescope (JWST, \citet{jwst2006, 2023PASP_Rigby}), particularly with its mid-IR ($\geq$3$\mu$m) capabilities, has unlocked the ability to acquire new and highly informative observations, building on the legacy of telescopes such as $Spitzer$ \citep{2004ApJS_Werner_spitzer} and $AKARI$ \citep{2007PASJ_Murakami_AKARI}. With JWST instruments such as NIRSpec \citep{2022A&A_NIRSpec_new} and MIRI \citep{jr:MIRI1, jr:MIRI4, jr:MIRI6, 2023PASP_MIRI_new} we can obtain spectra across 0.7 to 28 $\mu$m at medium resolution (approx 1,790-3750). This capability has been demonstrated as part of the \textit{The JWST Early Release Science Program for Direct Observations of Exoplanetary Systems} ($\#$1386, \citealt{2022PASP_Hinkley}), which acquired the first such spectrum of the planetary mass companion VHS 1256-1257 b (hereafter VHS 1256 b) \citep{2023ApJ_Miles}. VHS~1256 b's JWST data is comprised of NIRSpec and MIRI modes, which includes 12 orders across the full spectral range. This data comprises of  (1) NIRSpec\footnote{\url{https://jwst-docs.stsci.edu/jwst-near-infrared-spectrograph}} G140H/F100LP, G235H/F170LP and G395H/F290LP disperser and filter combinations producing a spectrum from 0.97-5.14$\mu$m at resolution $\sim$2700 and (2) MIRI\footnote{\url{https://jwst-docs.stsci.edu/jwst-mid-infrared-instrument/}} MRS across channels 1, 2 and 3 with varying resolution 1790-3750. The MIRI Channel 4 (18-28$\mu$m) SNR was too low to extract a spectrum and thus the data from C4 SHORT was binned to a single photometric data point \citep[see][]{2023ApJ_Miles}. These observations show a silicate cloud feature at 10$\mu$m, hints of CO$_{2}$ gas and simultaneous detections of both CO and CH$_{4}$ in the L and M band suggesting disequilibrium chemistry driven by strong vertical mixing.

Our preliminary modelling effort in \citet{2023ApJ_Miles} of the JWST data employed self-consistent modelling via PICASO 3.0 \citep{2023ApJ_Mukherjee}. This was followed by a much more extensive modelling investigation in \citet{2024ApJ_Petrus_2024} which employed the ForMoSA data fitting framework \citep{2020A&A_Petrus} to compare self consistent models ATMO \citep{2015_Tremblin}, Exo-REM \citep{Charnay_2018}, Sonora Diamondback \citep{2024_Morley}, BT-Settl \citep{2012_Allard} and DRIFT-PHOENIX \citep{2008_Helling}. \citet{2023ApJ_Gandhi} outlined a retrieval investigation using only the NIRSpec G395H data in pursuit of detecting carbon isotopes. Here we outline the first application of a retrieval framework to the full 1 -- 18 $\mu$m dataset as we aim to characterise the chemical and cloudy nature of the dynamic world VHS 1256 b. 

\section{VHS 1256 b}

\begin{figure}
\centering
\includegraphics[width=0.48\textwidth]{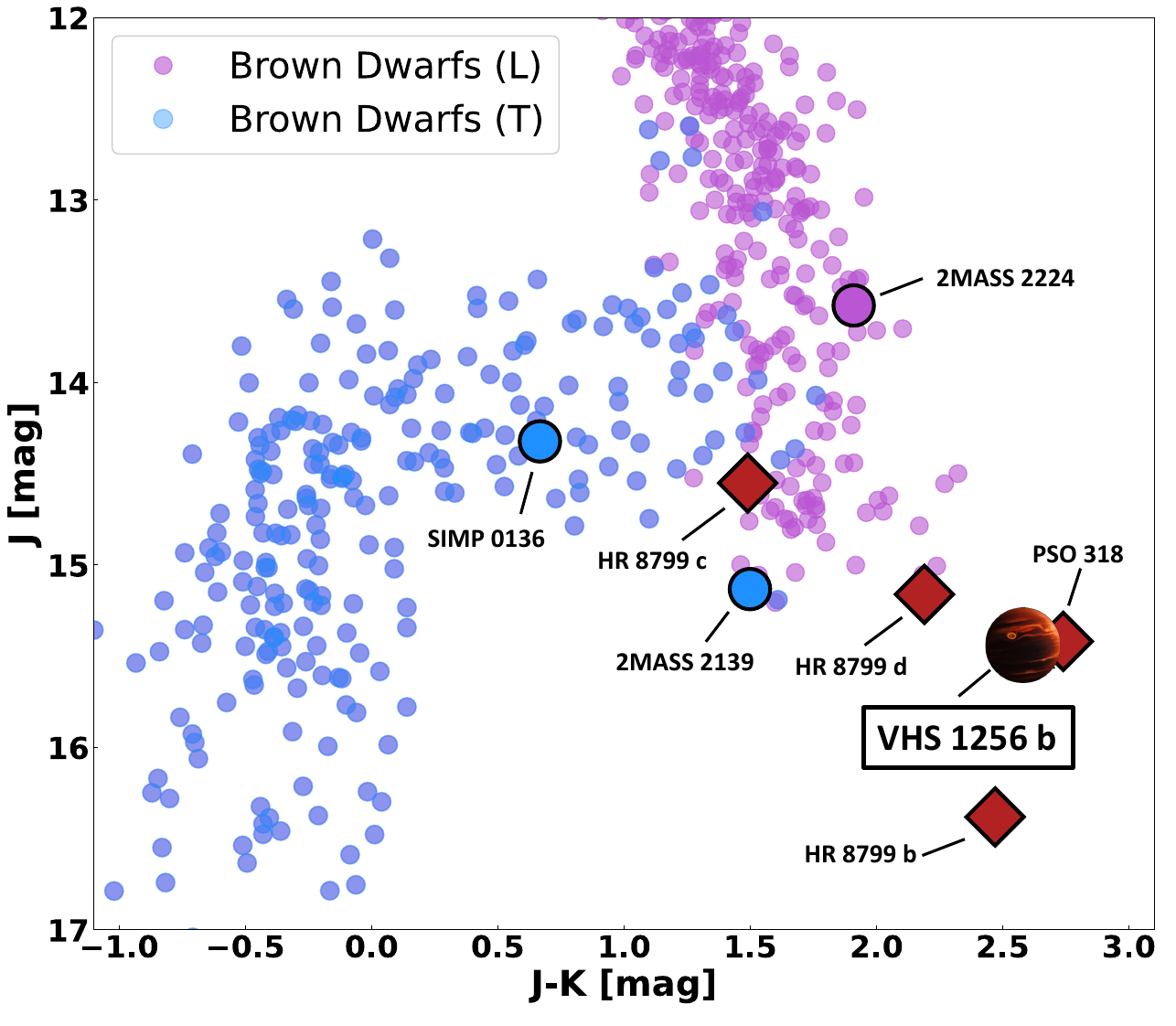}
\caption{Color magnitude diagram illustrating the brown dwarf color evolution across the L (magenta) and T (blue) sequences. We have illustrated VHS 1256 b's colors along with its spectral type counterparts including HR 8799 bcd and PSO 318. Also highlighted are 2MASSW J2224438-015852 (mid L), SIMP J01365662+0933473 and 2MASS J21392676+0220226 (early Ts) as they have been the subjects of the same $Brewster$ retrieval recipe (framework) that we employ in this study. Data used to compile this figure are from \citet{2020_Best_Catalog} catalogue which is a compilation of \citet{2012_Dupuy_liu, 2013_Dupuy_Kraus, 2016_Liu, 2018_Best} and \citet{2020_Best}.}
\label{fig:CMD}
\end{figure}

Discovered in 2015 \citep{Gauza_2015_VHS1256b}, VHS 1256 b is a young (140 $\pm$ 20 Myr, \citet{2023MNRAS_Dupuy}), red planetary mass ($\leq$ 25M$_{jup}$) companion orbiting an M dwarf binary at approximately 8'' or 179 $\pm$ 9 AU  \citep{2016_Stone, 2023A&A_Petrus}. It sits at a distance of 21.14 $\pm$ 0.22 pc \citep{2023A&A_GAIA_DR3, 2023MNRAS_Dupuy}. It is an L/T transition object, akin to objects such as the HR 8799 exoplanets \citep{Marois2008, Marois2010} and the free-floating object PSO J318.5338-22.8603 \citep[PSO 318][]{Liu_2013_PSOJ318}, positioning it perfectly to investigate the mechanisms and properties that drive the near-IR reddening of young low surface gravity brown dwarf and planetary mass companions [see Figure \ref{fig:CMD}.] 

VHS 1256 b has been the subject of several optical and near-IR spectral characterisation/observational campaigns \citep{Gauza_2015_VHS1256b,Miles_2018_PSOJ318,2022AJ_Hoch, 2023A&A_Petrus}. These mainly focused on the 0.5-2.5$\mu$m region which offers a window into H$_{2}$O and CO chemistry and abundances. \citet{Miles_2018_PSOJ318} obtained an L band spectrum attempting to detect and constrain CH$_{4}$. For young L/T objects, the ability to observe both CO and CH$_{4}$ features simultaneously is vital to probe states of chemical disequilibrium and its driving mechanisms, including vertical mixing and turbulence \citep{2015_Marley_Robinson}.

Before JWST, we had no observations spanning the mid-IR spectral signatures of silicate clouds for the young planetary mass, very red L type cohort of objects that includes VHS 1256 b (see Figure \ref{fig:CMD}). Now, for the first time, we have clear evidence of the broadband silicate feature (\cite{2023ApJ_Miles}, see Figure \ref{fig:intro_data_comparison}). For two decades, this silicate feature has been a well established characteristic of VHS 1256 b's brown dwarf cousins in the L type (2500-1200K) regime due to extensive observations using $Spitzer$ \citep{2005ApJ_Cushing, 2006ApJ_Cushing, 2009ApJ_Stephens}. More recently \citet{2022MNRAS_Suarez} produced a full re-reduction of these $Spitzer$ spectra. This work demonstrated that the strength of this silicate feature seems dependent on both surface gravity, the size of condensate grains present (driven by the surface gravity present) and inclination (viewing) angle \citep{2023MNRAS_Suarez_dust_grains, 2023ApJ_Suarez_Equatorial_Latitudes}.

Whilst photometric and spectral variability of substellar objects is common place \citep[e.g][]{2009ApJ_Artigau, 2012ApJ_Radigan, 2014ApJ_Burgasser, Biller_2015_PSOJ318, 2015ApJ_Metchev, 2016ApJ_Lew, 2018AJ_Manjavacas, 2018MNRAS_VosV, Vos_2019, 2019ApJ_Manjavacas, 2019A&A_Eriksson, 2020_Vos, 2022ApJ_Vos, 2024MNRAS_LiuPengyu, 2024arXiv_Biller}, VHS 1256 b is a standout object. It shows the highest level of spectrophotometric variability observed to date with several instruments. Using HST, \citet{2020_Bowler_vhs1256b} showed a $\sim25\%$ change in brightness and using Spitzer \citet{2020_Zhou_vhs1256b} showed a $\sim6\%$ shift in brightness. These studies were able to use these data to constrain a 21-24 hour rotation period, putting VHS 1256 b at the higher end of the rotation rate spread for variable substellar objects. \citet{2022AJ_Zhou} then used HST to monitor nearly 2 full rotations. Not only did \citet{2022AJ_Zhou} present a record setting change in brightness from the observations presented in \citet{2020_Bowler_vhs1256b} ($\sim38\%$ change in $J$ band brightness), but it also showed significant changes in the amplitude of variability from one rotation to the next. Objects like VHS 1256 b and its variable substellar kin have been theorised to have evolving atmospheres dominated by zonal waves, spots, or a mixture of both \citep[e.g][]{2013_Apai, 2015ApJ_Metchev, 2017Sci_Apai, 2021ApJ_Apai, 2022AJ_Zhou,2024ApJ_Fuda}. 

\section{Challenges to keep in mind: A complex analysis of a complex object with a complex dataset }

Our analysis centered around the application of the retrieval (inverse) method. This technique is data driven where a flexible model adjusts to fit observations without the physical constraints which cornerstone self consistent grid-model approaches. 

In this section we enumerate some of the obstacles that need to be overcome before we attempt a retrieval of JWST medium resolution spectroscopy (R$\sim$1790-3750) that is stitched from different instruments of a source with a complex atmosphere.


\subsection{A challenging object}
\label{sec:exotic_object}

VHS 1256 b sits in the middle of the exotic camp of very red, low surface gravity L-to-T type objects with temperature $\sim$1300-1000K \citep[e.g][]{2015_Marley_Robinson, 2016_Bowler, Biller_Bonnefoy_2018, 2023ASPC_Currie_Biller}. The near-IR SEDs of these L/T objects have very red colours (see Figure \ref{fig:CMD}), which has long been hypothesised to be driven by the presence of clouds \citep{1997_Burrows, AckermanMarley2001, 2002_Lodders_Fegley,2002_Marley, 2003_Tsuji_Nakajima, Marley_2010_clouds, 2015_Marley_Robinson, 2021_Gao, 2013_Faherty}. However, more recently an alternative mechanism of chemical instability (``fingering convection") which drives a non-adiabatic temperature-pressure gradient, and not clouds, has been been proposed as the key driver of the spectral signature of these objects \citep{2015_Tremblin, Tremblin_2016, Tremblin_2017, Tremblin_2019, 2020A&A_Tremblin}.  

After the \citet{2015_Tremblin, Tremblin_2016} theoretical predictions, several studies such as \citet{Molliere_2020}, \citet{2023MNRAS_Whiteford} and \citet{2024arXiv_Phillips}, explored $\sim$1-2.5$\mu$m spectra of similarly young and low surface gravity objects, including the HR 8799 planets and BD+60 1417B (or CWISER
J124332.12+600126.2) which have overlapping temperature estimates with VHS 1267 b. These studies directly encountered and demonstrated this model tension where a more isothermal (non-adiabatic) temperature-pressure gradient, resulting from flexible temperature-pressure parameterisations within the retrieval model, can be preferred over the inclusion of clouds. We note, however, these studies lacked mid-IR observations, specifically observations of the $\sim$10$\mu$m silicate feature.  VHS 1256 b is positioned at the crossroads of these competing model theories and, with its NIRSpec and MIRI, makes it an ideal laboratory for this debate and exploring this known model tension further. In considering how to retrieve the temperature-pressure profile for a source like VHS 1256 b, one must ensure that it is flexible enough to differentiate between atmospheric phenomena. 

\citet{Burningham_2021} showed how employing mid-IR (Subaru IRCS $L$ band  2.5-5.0$\mu$m and $Spitzer$ 5.2–14.2$\mu$m) data on top of near-IR (IRTF/SpeX 1-2.5$\mu$m) data made the retrieved temperature-pressure profile more adiabatic for 2MASSW J2224438-015852 (the archetypal L4 brown dwarf demonstrating a strong silicate feature). This mid-IR data provided the retrieval with the spectral signature of silicate clouds and acted to make the profile more consistent with those predicted by self-consistent models from \citet{2008_Saumon_Marley_Evolution_models} and \citet{2008ApJ_Helling_DRIFT_PHOENIX}.

VHS 1256 b is $\sim$400 Kelvin cooler than 2MASSW J2224438-015852, and has a lower surface gravity, putting it in a regime where we expect disequilibrium chemistry and intense atmospheric mixing \citep{Miles_2018_PSOJ318}.  Therefore we are in uncharted territory when attempting to confidently retrieve a temperature-pressure profile and cloudy structure that represents the truth for these very red L/T type objects when including mid-IR data with a clear $\sim$10$\mu$m silicate feature. As noted previously, VHS 1256 b is also the most spectrally variable planetary mass companion currently known \citep{2022AJ_Zhou}. The hypothesized underlying mechanism for this variability such as evolving bulk cloud coverage and/or convective atmospheric mixing \citep{2020A&A_Tremblin} only adds to the modelling challenge for this object.

\subsection{Order stitching and overlaps}
\label{sec:Order_stitching_and_overlaps}


The combined JWST observations of VHS 1256 b result in 11 overlapping zones (15 if MIRI Channel 4 is also considered) as highlighted in Figure  \ref{fig:intro_data_comparison} -- see Table 1 in \citet{jr:MIRI6} for further details.
Overlapping areas of spectral data from different instruments and stitching filters/gratings from the same instrument may present challenges within a retrieval framework. 
Notably, order overlap can occur both within a given instrument (e.g., the two overlaps in NIRSpec or the 8 overlaps in MIRI) and also via the addition of two instruments together (e.g., NIRSpec and MIRI). This is illustrated in Figure \ref{fig:intro_data_comparison}.
These consequences include: (1) There may be a systematic flux calibration offset between two orders which can introduce phantom spectral (molecular) features. A way to tackle this is to add a flux scaling parameter within the retrieval for each order or instrument \citep[e.g.][]{Burningham_2021, 2023ApJ_Vos, 2023MNRAS_Whiteford}. This comes with its own drawbacks such as the undesirable addition of a large amount of extra parameters within the retrieval framework, possibly inhibiting the convergence of the retrieval. (2) The retrieval may weight overlapping zones differently than non overlapping zones, when evaluating the BIC, due to there being more (x2) data points in these wavelength ranges. Depending on what spectral features are present in these overlaps, this may introduce biases within the retrieval, such as overfitting certain molecular features.
(3) Mismatches in the spectral signatures present in these spectral overlap zones, caused by imperfect and inconsistent data reductions, or spectral variability, could provide fitting challenges for the retrieval model. For example, if spectral variability is causing a mismatch, a single forward model is attempting to to fit the spectral signatures of VHS 1256 b at two differing rotational phases.

\subsection{Computational expense}

Retrieval analysis is a computationally intensive approach to model fitting compared to approaches such as grid model comparison. The run times of the forward model, which is the engine of the retrieval analysis, are dictated by several factors such as choice of chemistry framework, model spectra resolution,  cloudy vs. cloudless, and flexible vs. inflexible temperature-pressure profiles (2 vs $\geq$5 parameters). These choices can result in retrievals with 40+ modelled parameters (compared to less than five for a grid model comparison, as in \citet{2024ApJ_Petrus_2024}). Another crucial time determinant is the wavelength range and data resolutions that need modelling. For context, \citet{Burningham_2021} and \citet{2023ApJ_Vos} used the same retrieval framework as that employed in this work albeit over slightly shorter wavelength coverage (from 1-14.5$\mu$m instead of our 1-18$\mu$m in this case).  Between those two studies, they iterated through a list of model combinations that took approximately 3 weeks per retrieval when employing approximately 200 cores ($\sim$ 100,000 CPU hours) to achieve approximately 55 - 73 million MCMC model evaluations.  Consequently, retrieval studies require computational budgets on the order of millions of CPU hours.  That time only increases when wavelength coverage is extended, along with increasing spectral resolution, as forward models take longer to compute.



\begin{figure*}
\centering
\includegraphics[width=0.99\textwidth]{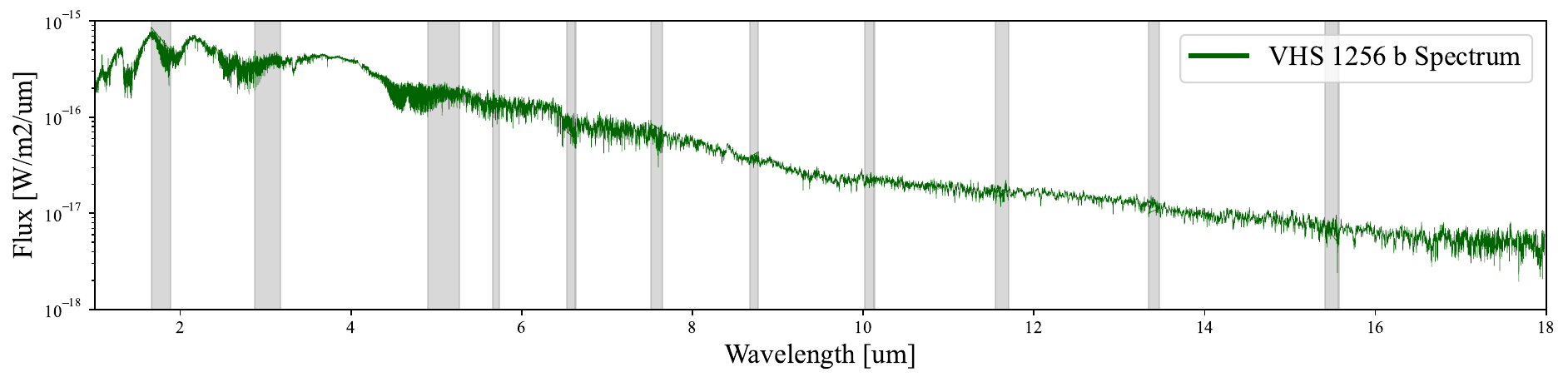}
\includegraphics[width=0.99\textwidth]{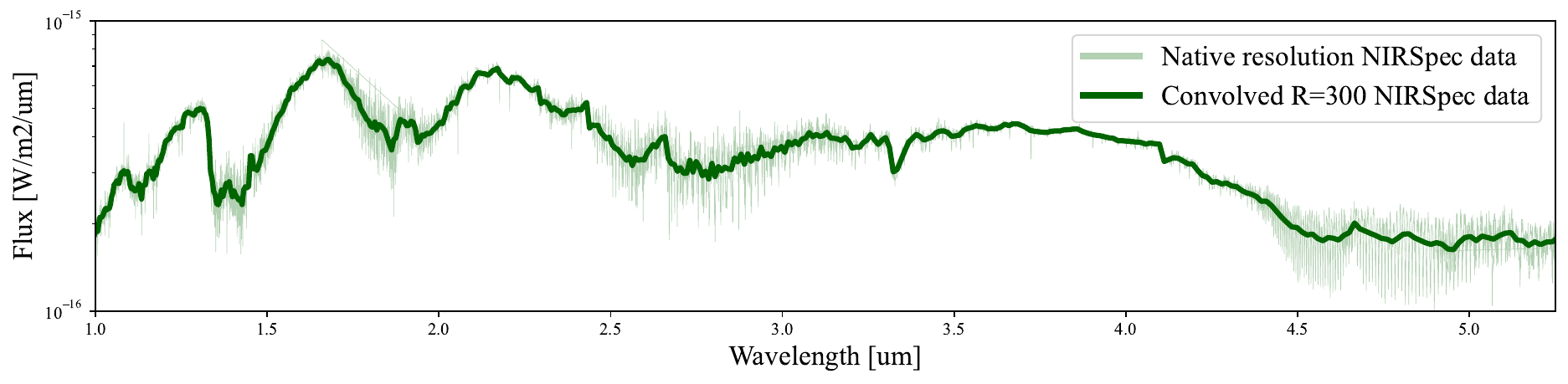}
\includegraphics[width=0.99\textwidth]{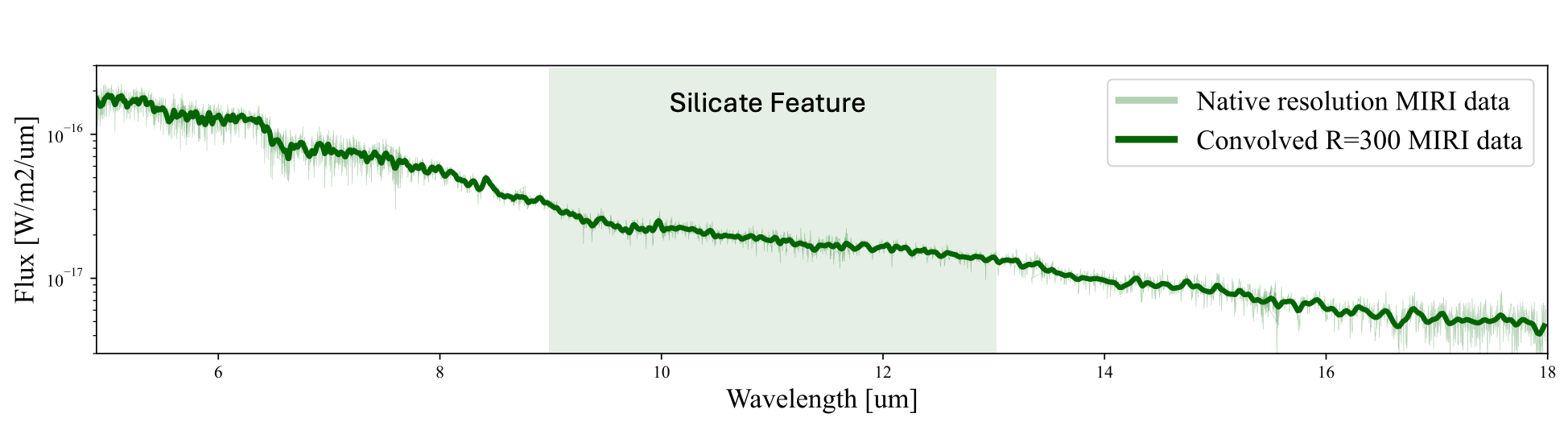}
\caption{Illustration of the VHS 1256 b data used in this study. (Top) Full ERS \#1386 program (V2) spectrum of VHS 1256 b with overlapping wavelengths between orders shown in grey. With 12 spectral orders populating the NIRSpec and MIRI data combination we have 11 overlaps present across the global spectrum. (Middle) NIRSpec data at its native resolution and the binned spectrum used in our retrieval analysis. (Bottom) MIRI data at its native resolution and the binned spectrum used in our retrieval analysis.} 
\label{fig:intro_data_comparison}
\end{figure*}

\section{RETRIEVAL RECIPE}

\label{sec:retrieval_recipe}

With all of the challenges described in section 3 in mind, we carefully formulated a retrieval recipe. For our analysis, we have used the $Brewster$\footnote{\url{https://github.com/substellar/brewster}} retrieval framework as outlined extensively in \citet{Burningham_2021} and \citet{2023ApJ_Vos}. In the following subsections we give an overview of this framework but refer the reader to \citet{Burningham2017}, \citet{Burningham_2021} and \citet{2023ApJ_Vos} for further insight into the inner workings of this retrieval code. $Brewster$ has been employed across a variety of L, T and Y type substellar objects \citep{Burningham2017, 2020_Gonzales, 2021ApJ_Gonzales, 2022ApJ_Gonzales, 2022ApJ_Calamari, 2023_Gaarn, 2023ApJ_Vos, 2023Natur_Barrado, 2024Natur_Faherty}. For this analysis we used the same prior boundaries outlined in Table 2 of \citet{Burningham_2021} except for surface gravity (log $g$) which was limited by a mass range of 1 -- 35 M$_{Jup}$.

\subsection{General setup}
$Brewster$ uses a 1-dimensional, two stream model which includes scattering \citep{1989_Toon, 1989_McKay}. We used Brewster's default model atmosphere with 64 layers distributed evenly in log space from 10$^{2.3}$ to 10$^{-4}$ bar.

We use the temperature-pressure profile outlined in \cite{jr:Mad&Seager2009} which has been employed regularly in a catalogue of substellar atmosphere retrieval investigations \citep[e.g.][]{Burningham2017,  2020_Gonzales, 2021ApJ_Gonzales,2023_Gaarn, Burningham_2021, 2023ApJ_Vos} as well as transiting exoplanet retrievals} \citep[e.g.][]{2018_Gandhi_Hydra, Pinhas_2018, MacDonald_Madhusudhan_2019}. This is a combination of exponential gradient curves, and hence does not permit sharp gradient discontinuities in the profile in the same fashion as more flexible geometrically interpolated temperature-pressure profile parameterizations \citep[e.g. $npoint$ profile,][]{TauREx3}. The \cite{jr:Mad&Seager2009} profile is split into three sections with pressure and temperature being related by:

\begin{align*}
P_{0} < P < P_{1}: P = P_{0}e^{\alpha_{1}(T-T_{0})^{1/2}} \ (Zone \ 1), \\
P_{1} < P < P_{3}: P = P_{2}e^{\alpha_{2}(T-T_{2})^{1/2}} \ (Zone \ 2), \\
P > P_{3}: T = T_{3} \ (Zone \ 3),
\end{align*}

\noindent where $T_{0}$ and $P_{0}$ are the top-of-atmosphere (10$^{-4}$ bar) temperature and pressure, and $T_{3}$ represents the isothermal temperature for atmospheric layers below $P_{3}$. Thermal inversions, while possible using this parameterization, are prohibited by setting $P_{2}$ = $P_{1}$, as inversions are not expected in the atmosphere of VHS 1256 b unlike in many hot Jupiters \citep[e.g.][]{2008ApJFortney_HotJups, 2019MNRAS_Gandhi_thermal_inversions}. With an inversion ruled out, and assuming continuity at zonal boundaries, we consider five free parameters within our analysis: $\alpha_{1}$, $\alpha_{2}$, $P_{1}$, $P_{3}$, and $T_{3}$. $P_{3}$, and $T_{3}$ act as an anchor point for the overall profile and can be set as values below the bottom-of-atmosphere (10$^{2.3}$ bar) pressure boundary of our modelled radiative transfer scheme. 

For our molecular species we use the so-called ``free chemistry" approach (no chemical equilibrium constraints) via the use of molecular abundances with constant vertical mixing ratios with each ratio sampled independently. Within our retrievals in this study we included an extensive list of molecular species which are predicted to be present and play a role in shaping the spectrum of an $\sim$1200K object such as VHS 1256 b. This included H$_{2}$O, CO, CO$_{2}$, CH$_{4}$, NH$_{3}$, TiO, VO, CrH, FeH, H$_{2}$S, SiO, Na, and K from \citet{2008ApJS_Freedman, 2014ApJS_Freedman}. These opacities have a native resolution of R = 10,000 where the model is initially computed at this resolution before being convolved to the resolution set by a tailored instrument mode. This convolved spectrum is then binned to the wavelength grid of the input data.

\vspace{0.2cm}

\subsection{Resolution, error bars and scaling}

\subsubsection{Resolution}

With the consideration of retrieval run times and testing retrievals on such extensive high resolution data for the first time, we elected to analyze a convolved version of the spectrum to a uniform resolution of R=300, using Brewster's convolution approach, which is then binned to a wavelength grid to mirror oversampling at 3 pixels per resolution element, approximating NIRSpec and MIRI's oversampling rate. This allowed for the more efficient running of models as we can employ cross sections at R=10,000. In order to run retrievals on JWST data at the native MRS level we would require cross sections at R$\geq$30,000 in order to achieve a model resolution at least one order of magnitude higher than the data being fit. This would significantly increase model run times which -- as described in section 3.3 -- are already a bottleneck for this type of retrieval analysis.  

\subsubsection{Error bars}

\begin{figure}
\centering
\includegraphics[width=.45\textwidth]{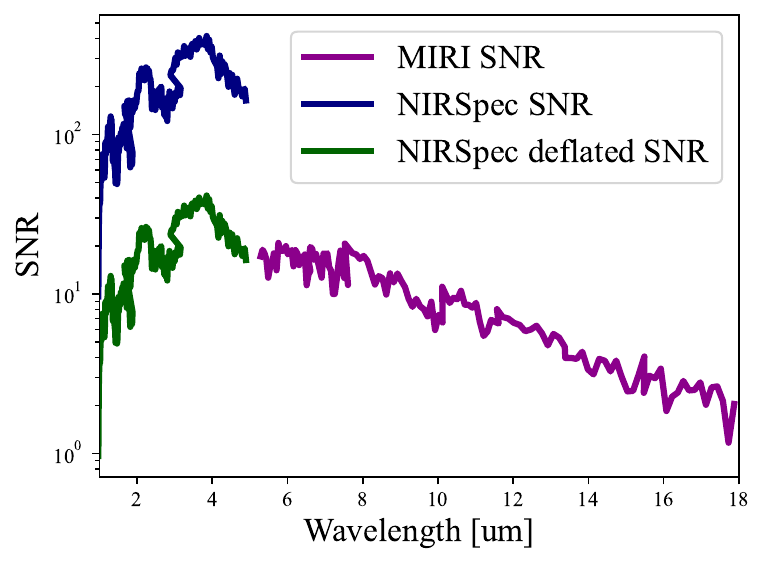}
\caption{Illustration of VHS 1256 b's NIRSpec and MIRI data SNRs using the V2 reduction from \citet{2023ApJ_Miles}. The NIRSpec data have an SNR approximately an order of magnitude higher than that of MIRI data. In our main analysis we deflated the NIRSpec SNR by an order of magnitude to more evenly match that of MIRI for a more equal instrument weighting in the retrievals.}
\label{fig:data_SNR}
\end{figure}

When the VHS 1256 b data reduction evolved from Version 1 to Version 2, the NIRSpec portion of the data saw a dramatic increase in signal to noise ratio (SNR). Early in our retrieval analysis with the V2 reduction, we encountered clear issues with the weighting of the NIRSpec data vs. the weighting of the MIRI data. These early V2 retrievals clearly focused on fitting (likely overfitting) the NIRSpec data and almost entirely ignored the more subtle features  present in the MIRI data, particularly the silicate feature at 8.5-10 microns. As a result, for our retrieval analysis comparison of cloud combinations we increased the error bars (deflated the SNR) of the NIRSpec V2 data by a factor of 10 as shown in Figure \ref{fig:data_SNR}. This resulted in NIRSpec and MIRI data having a much more comparable SNR.

We found that increasing the error bars of the NIRSpec data to more closely match that of MIRI (see Figure \ref{fig:data_SNR}) allowed for an improved model fit across the MIRI dataset, particularly the silicate feature near 10 $\mu$m. 

As mentioned previously we convolved the JWST data from their native resolutions to a constant R=300. The resulting errors for this resolution were produced using the same convolution. We adopted this cautious approach, instead of binning, to avoid overly precise errors bars when testing our retrieval framework against such a novel dataset. For example, using a simple $Error\sim\sqrt{n}$ assumption of uncorrelated data, binning the NIRSpec/G395H (V2) data from R=2,700 to R=300 would have resulted in an SNR $> 1000$ for a significant portion of the wavelength coverage. We also note that we employed the commonly used error tolerance b factor \citep{2013_Foreman-Mackey, Line_2015} for each of the NIRSpec orders (G140H, G235H and G395H gratings) but only one for the entire MIRI dataset. This resulted in a total of 4 tolerance b factors within our retrievals. 

\subsubsection{Scaling}

As we alluded to in subsection 3.2, for the JWST VHS 1256 b, the number of orders that make up the spectrum presents a challenge (see Figure \ref{fig:data_SNR}). Our treatment of scaling factors evolved during our analysis. Initially we tested (1) no scaling factor and (2) a scaling factor per order (12 in total) with flat priors between 0.5 and 1.5. Both treatments were problematic in terms of model dimensionality and led to behaviors such as the retrieval acting to essentially remove the broadband silicate feature (present in MIRI) by scaling the MIRI orders spanning this broadband silicate feature to mimic a more blackbody and featureless slope. Therefore, we settled on a middle ground approach where the NIRSpec/G140H order acts as the model anchor with independent scaling factors for NIRSpec/G235H and NIRSpec/G395H with the MIRI data then only having a single scaling factor across all 3 channels of data.

\subsection{Model Selection}

In order to differentiate between the goodness-of-fit for different retrievals we use the Bayesian Information Criterion (BIC). The preferred model is used as the reference where $\Delta$ BIC increases with degrading goodness-of-fit. The BIC is given by:

\begin{equation}
BIC = k \ln(n) - 2\ln(L)
\end{equation}

\noindent where $k$ is the number of parameters, $n$ is the number of data points and $L$ is the likelihood. \citet{Kass_Raftery_1995} define BIC magnitudes that define different confidence levels for models:  

\begin{align*}
\mathrm{0 < BIC < 2:} \mathrm{no \; preference} \\
\mathrm{2 < BIC < 6:} \mathrm{indicative} \\
\mathrm{6 < BIC < 10:} \mathrm{strongly \; indicative} \\
\mathrm{BIC > 10:} \mathrm{very \; strongly \; indicative} \\
\end{align*}

\subsection{Cloud modelling approach}

A major goal of this work is to parametrize the clouds on VHS 1256 b. This is a critical aspect of the global characterization effort of VHS 1256 b (and similar objects) as the color differences between high and low surface gravity objects at the L/T transition (see Figure \ref{fig:CMD}) has been postulated to be due to sinking and/or stalled cloud breakup \citep{AckermanMarley2001,2002_Burgasser_T_Dwarfs,2003_Tsuji_Nakajima,2009ApJ_Stephens,Marley_2010_clouds,2021_Gao}. Therefore, VHS 1256 b is an ideal candidate to test these predictions using $Brewster$'s extensive cloud modelling capabilities (see \citet{Burningham_2021} for a thorough description).

\begin{figure*}
\centering
\includegraphics[width=0.9\textwidth]
{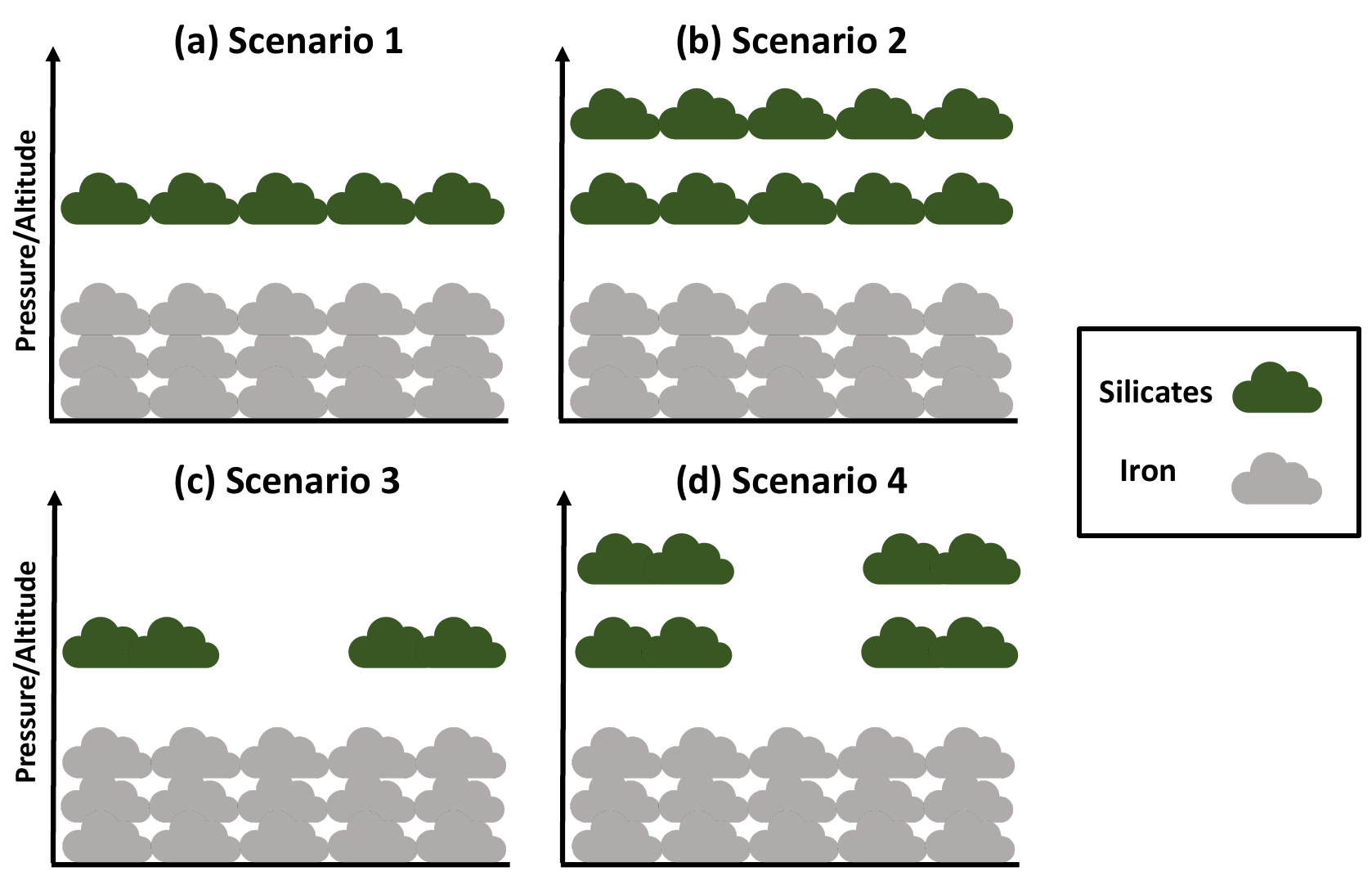}
\caption{This figure outlines the various cloud scenarios we investigated for VHS 1256 b. (a) illustrates scenario 1 which combined an iron deck with a uniform silicate slab cloud. (b) illustrates scenario 2 which combined an iron deck cloud with 2 uniform silicate slab clouds. (c) illustrates scenario 3 which combined an iron deck clouds with a patchy silicate slab cloud. (d) illustrates scenario 4 which combined an iron deck clouds with 2 patchy silicate slab clouds.}
\label{fig:cloud_scenarios}
\end{figure*}

The models we explored within this study were selected using 3 motivating factors: (1) The methodolgy and results of \citet{Burningham_2021} and \citet{2023ApJ_Vos}, (2) the variable nature of VHS 1256 b and (3) the computational expense of running an extensive array of retrievals. With these considerations in mind, we employed a similar cloud investigation framework as that in \citet{Burningham_2021} and \citet{2023ApJ_Vos}, which included objects which bookend the temperature of VHS 1256 b. These cloudy combinations, illustrated in Figure \ref{fig:cloud_scenarios} and outlined in Table \ref{tab:retrievalBICs}, are akin to the preferred models from the \citet{2023ApJ_Vos} studies but we note that the tally of model combinations tested is not as extensive. 

\citet{Burningham_2021} and \citet{2023ApJ_Vos} employed so-called ``deck" and ``slab" parameterisations to map the cloud structures present \citep{Burningham2017}. In the ``slab'' framework we retrieve both the cloud base and cloud top pressures of the cloud layer and the optical depth of the cloud. In the ``deck'' we retrieve only the cloud top pressure, with the the cloud becoming and remaining optically thick (opaque) to all deeper atmospheric layers. 
We also retrieve a decay height parameter that defines how quickly the cloud opacity decays above the top pressure, and ramps up below it.
This is defined in ${d\log P}$, and the height sets the pressure scale over which the opacity decreases by a factor $1/e$. 
For approach values that approach or exceed the scale height of the atmosphere, the actual optical depth asymptotes to the scale height of the atmosphere such that the maximum decay scale in $d\log P = \log(1/e)$. 

Generally the retrievals from \citet{Burningham_2021} and \citet{2023ApJ_Vos} showed a preference for an iron cloud deck deep ($\geq$1$\mu$m) in the atmosphere with a thin silicate cloud layer, or two, in or above the photosphere zone ($\sim$1--10 bar). In \citet{2023ApJ_Vos}, patchy clouds were the preferred model fits to the variable objects SIMP 0136 and 2M2139 \citep{2009ApJ_Artigau, 2012ApJ_Radigan}, something we expect to be present for VHS 1256 b based on its variability. Therefore, we took our model guidance from these studies when compiling the list of retrieval cloud combinations we wished to test. 
We investigated 4 cloud scenarios: (a) A deck and slab, (b) a deck and 2 slabs, and for both (c) and (d), the former two with the added dimension of cloud patchiness. In the patchy model we follow the commonly used approach  \citep[e.g.,][]{Marley_2010_clouds, 2023ApJ_Vos, 2023MNRAS_Whiteford} of linearly combining the models with and without the cloud(s) of interest using

\begin{equation}
F_{Tot} = C_{frac}\cdot F_{Cloudy} + (1-C_{frac})\cdot F_{Clear}
\end{equation}

\noindent where $F_{Tot}$ is the total flux, $F_{Clear}$ is the flux from a model without the cloud(s), $F_{Cloudy}$ is the flux from a model with the cloud(s) and $C_{frac}$ is the fraction of cloud coverage. We note that the clear and cloudy models share the same non-cloud properties.

In our retrieval setups the deck is set as Iron (Fe) with the slab or slabs set as some form or combination of amorphous silicate species:  (1) Forsterite (Mg$_2$SiO$_4$, \cite{1996ApJ_Scott_Duley_silicate_data}), (2) Enstatite (MgSiO$_3$, \cite{1996ApJ_Scott_Duley_silicate_data}) and (3) Quartz (Silicon Dioxide SiO$_2$, \cite{2015A&A_Wakeford_Sing}) -- see \ref{tab:retrievalBICs} for the list of specific combinations tested. For the particle size distribution for each of these species we employed a Hansen distribution \citep{Hansen_PSD_1971} where the distribution of n particles with radius r is defined by:

\begin{equation}
n(r) \propto r^\frac{1-3b}{b} e ^{-\frac{r}{ab}}
\end{equation}

with a and b define the effective radius and spread of the distribution and are the parameters retrieved.

\section{RESULTS: ATMOSPHERIC INFERENCES OF VHS 1256 b}
\label{sec:results}

\begin{sidewaystable}
\centering
\begin{tabular}{c|cccccccccccc}
\hline
\hline
& H$_{2}$O & CO & CO$_{2}$ & CH$_{4}$ & NH$_{3}$ & Log(g) & R$_{Jup}$ & M$_{Jup}$ & C/O & [M/H] & T$_{eff}$ \\ 
\hline
\hline
This work & -3.50$^{+0.05}_{-0.06}$ & -3.00$^{+0.07}_{-0.11}$ & -6.73$^{+0.07}_{-0.13}$ & -5.44$^{+0.05}_{-0.06}$ & -5.86$^{+0.23}_{-0.27}$ & *4.68$^{+0.03}_{-0.12}$ & 1.29$^{+0.02}_{-0.03}$ & *32.09$^{+2.16}_{-8.22}$ & **0.63$^{+0.01}_{-0.02}$ & 0.30$^{+0.06}_{-0.10}$ & 1155$^{+11}_{-9}$\\

\hline

BT-SETTL, P24 & - & - & - & - & - & 3.62$\pm$0.60 & 0.95$\pm$0.16 & $<$ 20.90 & - & - & 1560$\pm$203\\
DRIFT-PHOENIX. P24   & - & - & - & - & - & 4.54$\pm$1.13 & 0.93$\pm$0.15 & $<$ 249.42 & - & -0.36$\pm$0.26 & 1540$\pm$286 \\
ATMO, P24  & - & - & - & - & - & 3.59$\pm$0.53 & 1.09$\pm$0.23 & 2.24$\pm$1.96 &  0.46$\pm$0.16 & -0.32$\pm$0.28 & 1318$\pm$166 \\
Exo-REM, P24  & - & - & - & - & - & 3.70$\pm$0.34 & 1.39$\pm$0.28 & 7.87$\pm$6.33  & 0.38$\pm$0.20 & 0.03$\pm$0.17 & 1153$\pm$152 \\
SONORA, P24  & - & - & - & - & - & 4.5$\pm$0.63 & 1.12$\pm$0.16 & $<$ 94.40  & - & - & 1349$\pm$208 \\
\hline
\hline

\end{tabular}
\caption{Table of retrieved parameter values from our analysis along with the self consistent model values derived in the \citet{2024ApJ_Petrus_2024} study, denoted as P24. * Parameter converged to upper limit of the permitted parameter sampling space. ** We corrected for oxygen depletion using approach outlined in \citet{2024ApJ_Calamari}.} 
\label{tab:bulk_results_comparison} 
\end{sidewaystable}

Here we outline the retrieval results, discussing the global results of our bulk, molecular chemistry and cloud property inferences. For the results outlined here, as discussed in Section \ref{sec:retrieval_recipe}, we used data where the NIRSpec error bars were inflated. Please see the appendix where we explored the impact of inflating errors and/or just performing retrievals using only the NIRSpec data.

\begin{sidewaystable}
\centering
\begin{tabular}{ccccc}
\hline
Cloud 1 & Cloud 2 & Cloud 3  & Retrieval Rank & $\Delta$BIC  \\
\hline

Patchy Mg2SiO4 Slab & Patchy MgSiO3 Slab & Fe Deck & 1 (Most preferred) & 803.20 \\
Mg2SiO4 Slab & MgSiO3 Slab & Fe Deck & 2 & 506.24 \\ 

Mg2SiO3 Slab & SiO2 Slab & Fe Deck & 3 & 429.43 \\ 

MgSiO3 Slab & Fe Deck & - & 4 & 217.99 \\ 

Mg2SiO4 Slab & SiO2 Slab & Fe Deck & 5 & 52.62 \\ 

Mg2SiO4 Slab & Fe Deck & - & 6 (Least preferred) & 0 \\ 

\hline
\end{tabular}
\caption{Table outlining the Cloud models explored for VHS 1256 b in our retrievals along with their ranks with accompanying $\Delta$BIC and the total numbers of parameters in each retrieval.}
\label{tab:retrievalBICs} 
\end{sidewaystable}

\begin{figure*}
\centering
\includegraphics[width=0.93\textwidth]{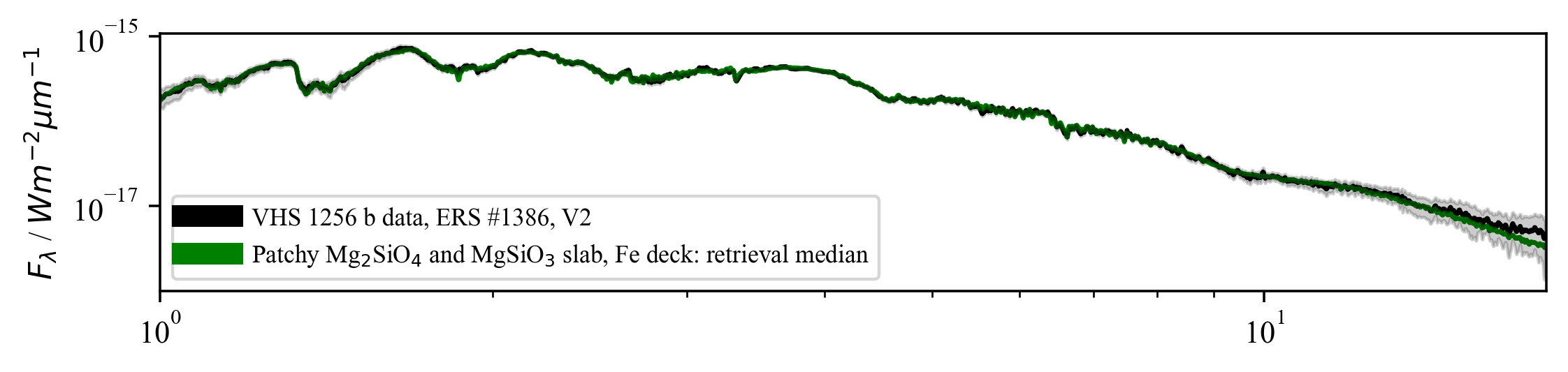}
\includegraphics[width=0.9\textwidth]
{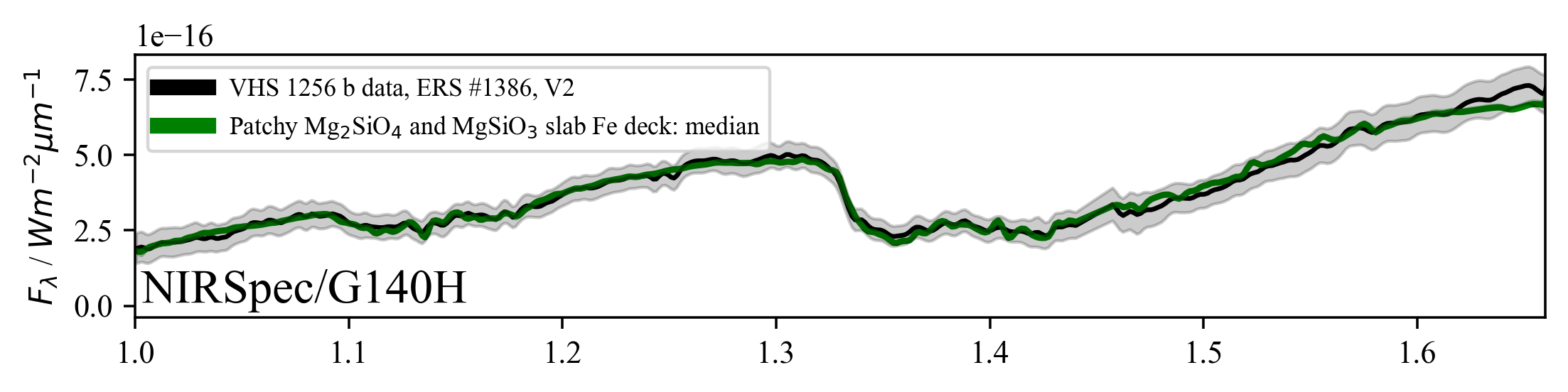}
\includegraphics[width=0.9\textwidth]
{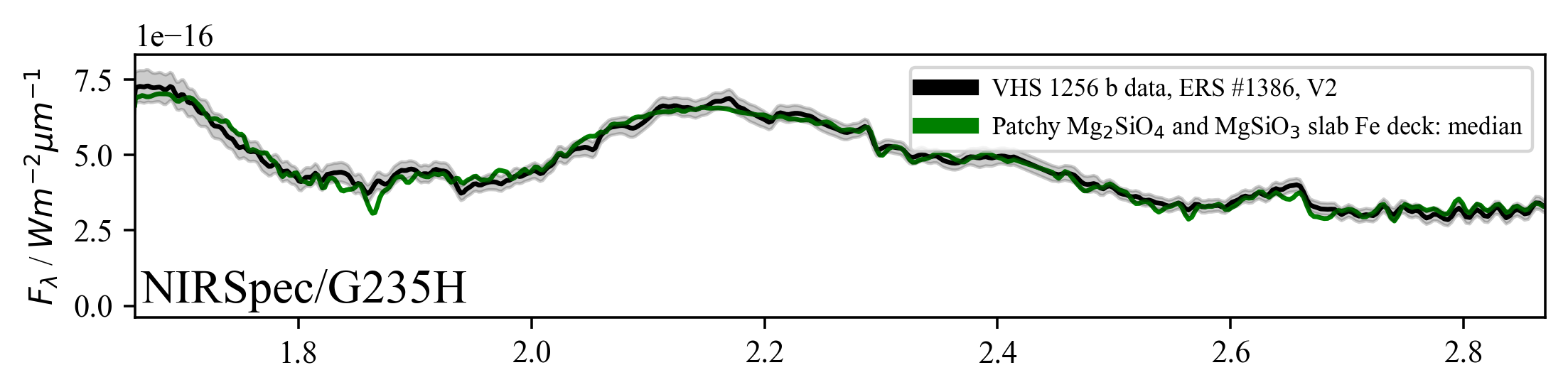}
\includegraphics[width=0.9\textwidth]
{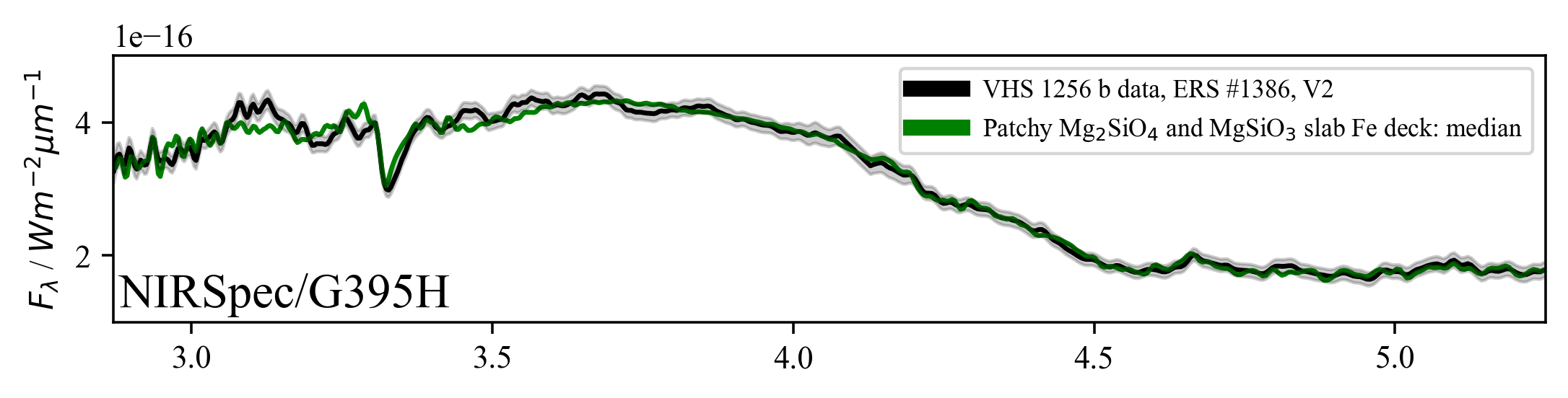}
\includegraphics[width=0.93\textwidth]
{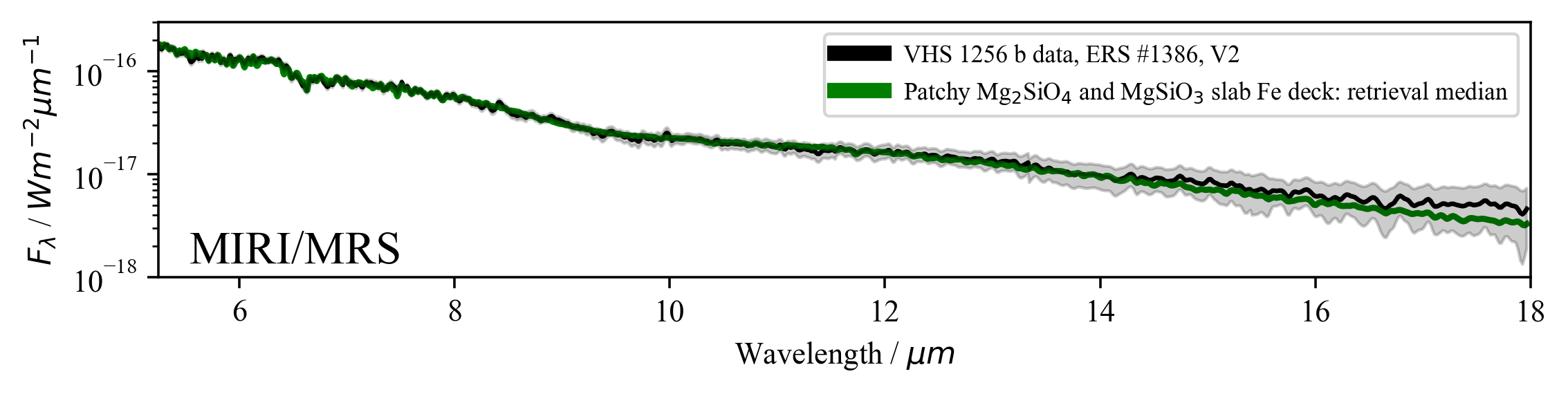}
\caption{Best fits from our retrieval analysis of VHS 1256 b. We show the global fit along with a zoomed illustration to NIRSpec/G140H, NIRSpec/G235H, NIRSpec/G395H and the MIRI Channels 1, 2 and 3 data. We note that this fit is to the spectrum when binned to resolution R=300 and with NIRSpec error bars inflated by an order of magnitude. See section \ref{sec:retrieval_recipe} for more details.}
\label{fig:spectral_fits}
\end{figure*}

\begin{figure*}
\centering
\includegraphics[width=0.9\textwidth]
{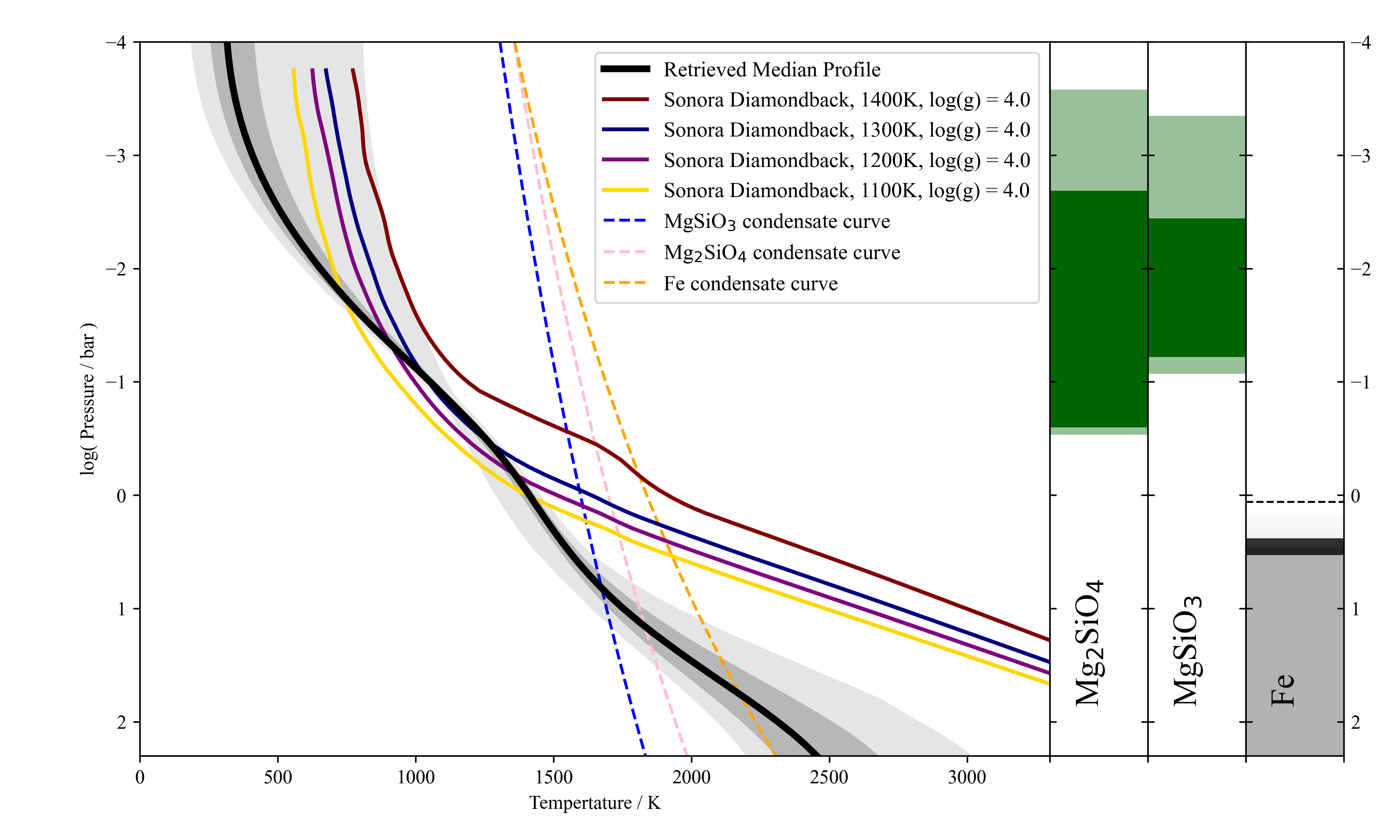}
\caption{(Left) The retrieved temperature pressure profile. For comparison we have also illustrated several temperature pressure profiles from self-consistent grid models Sonora Diamondback with f$_{sed}$=3. Condensate curves of relevant species explored in this work are also indicated. We also show the vertical (pressure) extent of the cloud included in the model. (Right) Green indicates the distribution in pressure of the silicate clouds, with shading showing the 1$\sigma$ uncertainty. The uniform gray indicates the distribution in pressure of the optically thick part of the iron deck cloud, with the black region showing the  1$\sigma$ uncertainty the retrieved pressure of the top of this cloud}. 
\label{fig:bestfitTP}
\end{figure*}

\begin{figure*}
\centering
\includegraphics[width=0.8\textwidth]{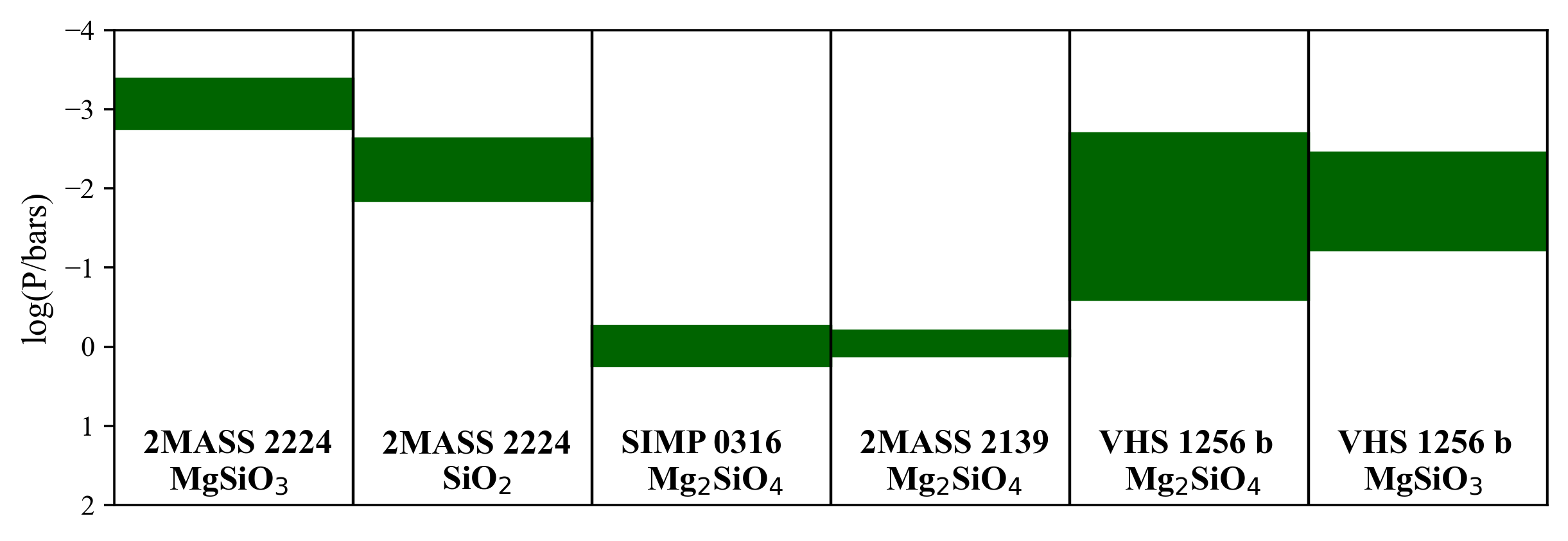}
\includegraphics[width=0.8\textwidth]{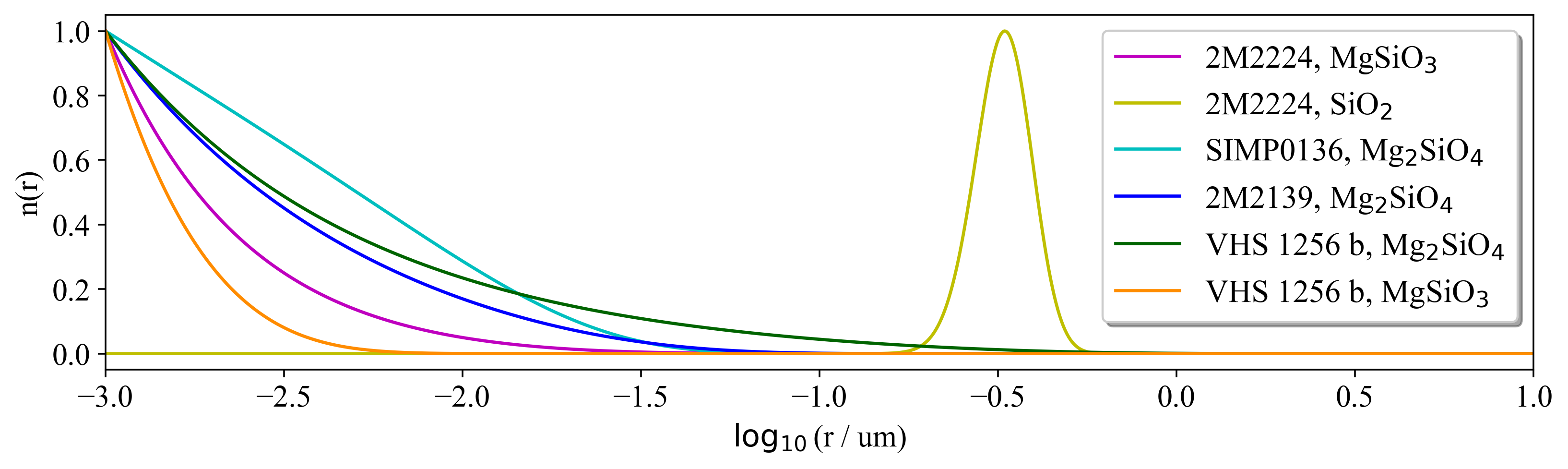}
\includegraphics[width=0.8\textwidth]{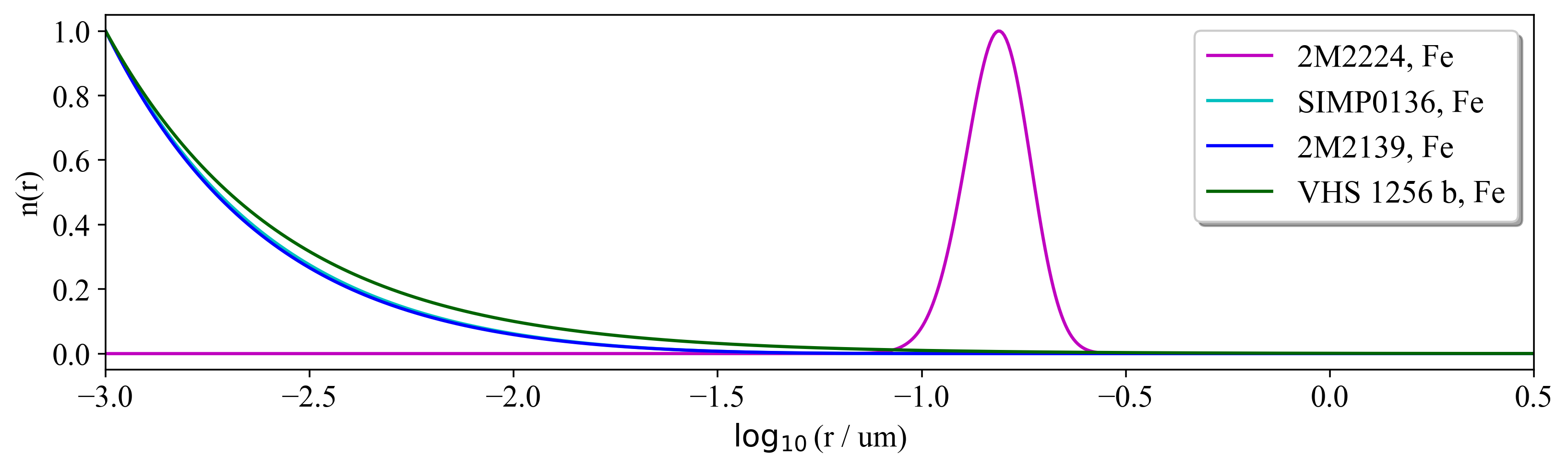}
\caption{Top: Comparison of retrieved cloud position for VHS 1256 b and objects from \citet{Burningham_2021} and \citet{2023ApJ_Vos}. Middle and Bottom: Comparison of retrieved particle size distributions for silicate and iron clouds for VHS 1256 b and objects from \citet{Burningham_2021} and \citet{2023ApJ_Vos}.}
\label{fig:cloud_property_illusration}
\end{figure*}

\begin{sidewaystable}
\begin{tabular}{c|ccccccc}
\hline
\hline
Object & Species & Coverage & $\tau_{cloud}$ (1$\mu$m) & log(P) [bar] & dlog(P) [bar] & radius, a & spread, b \\
\hline
\hline
& & & & Slab clouds & & &  \\
\hline
\hline
VHS 1256 b (This Work) & Mg2SiO4 & 0.985$^{+0.012}_{-0.022}$ & 5.21$^{+0.90}_{-1.83}$ & -0.60$^{+0.06}_{-0.07}$ & 2.09$^{+0.89}_{-0.88}$ & -0.05$^{+0.06}_{-0.05}$ & 0.42$^{+0.14}_{-0.13}$   \\
VHS 1256 b (This Work) & MgSiO3  & 0.985$^{+0.012}_{-0.022}$ & -2.36$^{+0.13}_{-0.13} (log)$ & -1.22$^{+0.15}_{-0.26}$ & 1.23$^{+0.90}_{-0.62}$ & -2.50$^{+0.31}_{-0.31}$ & 0.50$^{+0.32}_{-0.35}$   \\
2MASSW J2224 (B21) & MgSiO3   & N/A & 0.30$^{+0.32}_{-0.18}$ & -2.76$^{0.66}_{-0.44}$ & 0.62$^{+0.67}_{-0.44}$ & -1.41$^{+0.18}_{-0.17}$ & 0.53$^{+0.33}_{-0.36}$   \\
2MASSW J2224 (B21) & SiO2  & N/A & 4.54$^{+0.58}_{-0.58}$ & -1.85$^{+0.63}_{-0.99}$ & 0.78$^{+1.07}_{-0.58}$ & -0.44$^{+0.04}_{-0.20}$ & 0.03$^{+0.18}_{-0.01}$   \\
SIMP J0136 (V23)& Mg2SiO4  & 0.70$^{+0.03}_{-0.04}$ & 12.03$^{+1.03}_{-0.85}$ & 0.24$^{+0.06}_{-0.05}$ & 0.50$^{+0.09}_{-0.1}$ & -1.47$^{+0.14}_{-0.14}$ & 0.36$^{+0.33}_{-0.25}$   \\
2MASS J2139 (V23)& Mg2SiO4  & 0.83$^{+0.06}_{-0.06}$ & 8.17$^{+1.54}_{-1.38}$ & 0.12$^{+0.08}_{-0.09}$ & 0.32$^{+0.20}_{-0.15}$ & -1.21$^{+0.18}_{-0.19}$ & 0.42$^{+0.34}_{-0.29}$   \\
\hline
\hline
& & & &  Deck clouds & &  \\
\hline
\hline
VHS 1256 b (This Work) & Fe  & N/A & N/A & 0.46$^{+0.07}_{-0.08}$ & 3.68$^{+2.12}_{-2.00}$ & 1.73$^{+0.86}_{-0.98}$ & 0.50$^{+0.32}_{-0.32}$   \\
2MASSW J2224 (B21) & Fe  & N/A &  N/A & 0.95$^{+0.06}_{-0.06}$ & 4.2$^{+1.95}_{-2.25}$ & 0.77$^{+0.05}_{-0.06}$ & 0.03$^{+0.04}_{-0.01}$   \\
SIMP J0136 (V23)& Fe  & N/A & N/A & 0.90$^{+0.04}_{-0.05}$ & 0.06$^{+0.03}_{-0.02}$ & -1.35$^{+0.25}_{-0.31}$ & 0.51$^{+0.33}_{-0.33}$   \\
2MASS J2139 (V23) & Fe  & N/A & N/A & 0.86$^{+0.09}_{-0.07}$ & 0.03$^{+0.03}_{-0.02}$ & -1.31$^{+0.45}_{-0.66}$ & 0.52$^{+0.34}_{-0.35}$   \\
\hline
\hline
\end{tabular}
\caption{Retrieved cloud properties from this work along with those from \citealt{Burningham_2021} (B21) and \citealt{2023ApJ_Vos} (V23).} 
\label{tab:cloud_results_comparison} 
\end{sidewaystable}

\begin{figure*}
\centering
\includegraphics[width=0.99\textwidth]
{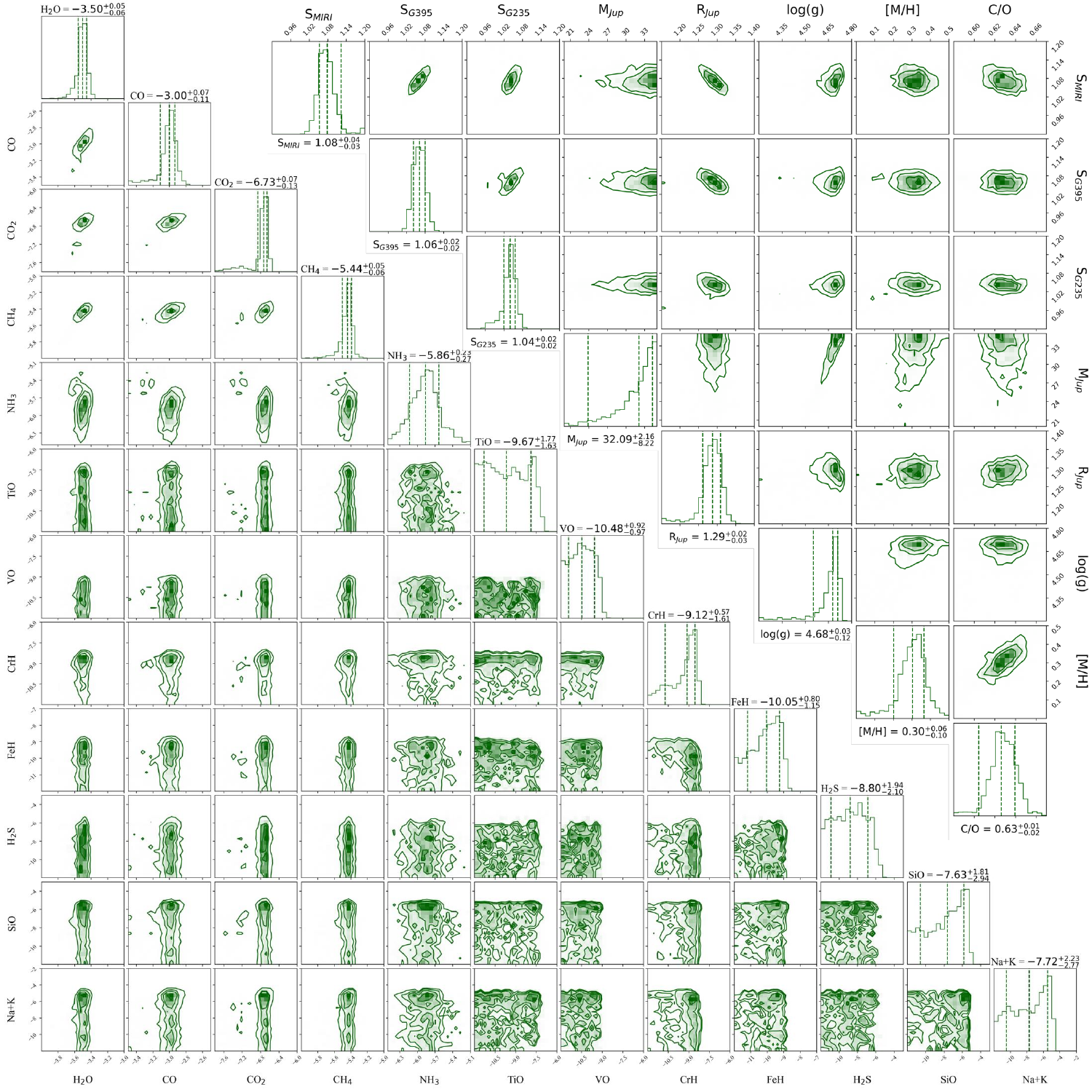}
\caption{Corner plots of retrieved values for the retrieval preferred model fit to VHS 1256 b. Bottom left: retrieved posteriors of molecular chemistry. Top right: retrieved posteriors of bulk parameters and spectral scaling factors. We note that Radius (R), Mass (M), C/O and [M/H] are inferred and not directly retrieved properties.}
\label{fig:chem_and_bulk_corner_plot}
\end{figure*}

\subsection{The best fitting cloud model: patchy forsterite and enstatite}

In Table \ref{tab:retrievalBICs}, we outline the various cloud models tested for VHS 1256 b. The preferred model consisted of a forsterite slab cloud, an Enstatite slab cloud and an iron deck cloud, producing the spectral fits shown in Figure \ref{fig:spectral_fits}. The preference for forsterite to be included aligns with the results of \citet{2023ApJ_Vos}. 
The preference for an additional silicate slab also aligns with  \citet{Burningham_2021}, who found the best fit retrieval model for 2M2224 combined enstatite with Silicon Dioxide.

Given the patchy cloud preference results of \citet{2023ApJ_Vos}, and the fact that VHS 1256 b is highly variable in the near-infrared, we expect that the atmosphere of VHS 1256 b is non-uniform and hence that our retrieval framework may prefer a patchy model as well (see Section \ref{sec:retrieval_recipe} for details). With this in mind, we ran models that include patchiness, as shown in Table \ref{tab:retrievalBICs}. Our results show a significant preference for the patchy silicate clouds ($\Delta$BIC = 297; scenario 4 in Figure \ref{fig:cloud_scenarios}). We constrain the cloud coverage fraction to be 0.985$^{+0.012}_{-0.022}$ (or 0.0155$\%$ cloud free). This is much smaller than the cloud coverage fractions of 0.7$^{+0.03}_{0.02}$ and 0.83$^{+0.06}_{0.06}$ retrieved for SIMP 0136 and 2M2139 respectively from \citet{2023ApJ_Vos}. However, a very high cloud coverage fraction, and by extension a very low cloud-free coverage, is approximately consistent with the modeling from \citet{2020_Bowler_vhs1256b} who found that varying the cloud free fraction from 0.00775 to 0.01225 best explained VHS 1256 b's spectral variability observations.

\subsection{The cloud properties}

Figure \ref{fig:bestfitTP} shows the thick forsterite and enstatite clouds in the atmosphere, between $\sim$10$^{-1}$ and $\sim$10$^{-3}$ bars -- see Figure \ref{fig:bestfitTP}. In contrast, \citet{2023ApJ_Vos} found the thinner forsterite clouds for SIMP0136 and 2M2139 were lower, positioned at approximately 1 bar. \citet{Burningham_2021} found the retrieved enstatite and silicon dioxide clouds to be at approximately 10$^{-2}$ to 10$^{-3}$ bar. These are illustrated in Figure \ref{fig:cloud_property_illusration}. VHS 1256 b's spectral type (L8) lies in the middle of the objects studied in \citet[]{Burningham_2021} (L4) and \citet{2023ApJ_Vos} (T2). 
Thus, if temperature is the primary driver of the atmospheric chemistry and cloud condition, and using condensate curves, one would expect the silicate cloud location for VHS 1256 b to be somewhere in-between the L4 and T2 objects. We see this is indeed the case for the bottom pressure of the silicate clouds but they appear much thicker (vertically extended in pressure space) than the cloud retrieved in \citet[]{Burningham_2021} and \citet{2023ApJ_Vos} -- see Figure \ref{fig:cloud_property_illusration}.
This could be explained by the low surface gravity combined with atmospheric mixing and/or dynamic turbulence \citep{2010A&A_Freytag, Barman_2011_HR8799b,Marley_2012_HR8799, 2022MNRAS_Tan} in VHS 1256 b's young atmosphere acts to propel clouds upwards making them more vertical distributed than objects such as 2M2224, SIMP0136 or 2M2139. However, definitively demonstrating such a relationship will require further testing across more objects and the incorporation of self-consistent cloud modelling.

\begin{figure}
\centering
\includegraphics[width=0.45\textwidth]{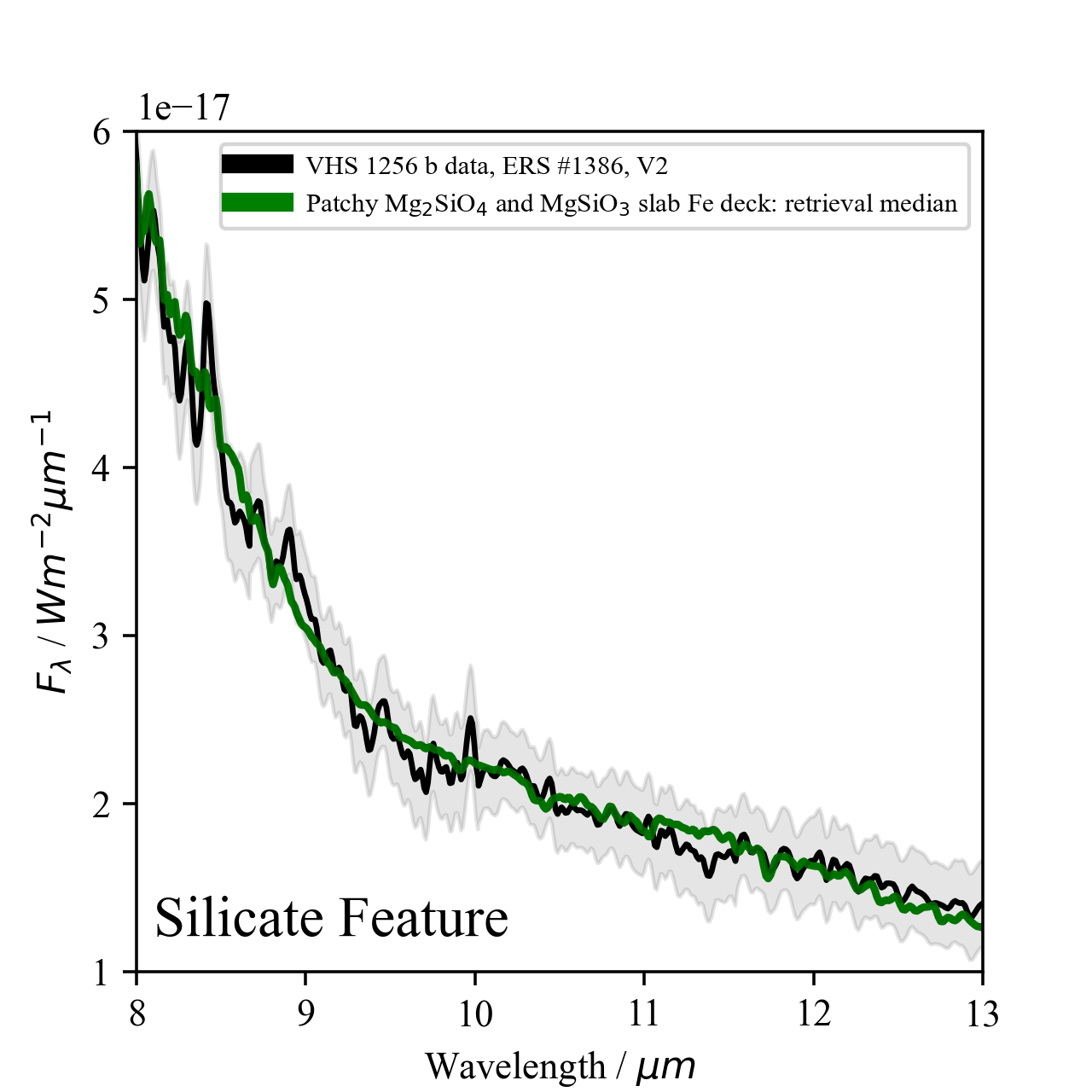}
\caption{Our preferred retrieval's median fit of VHS 1256 b's silicate feature via a combination of forsterite and enstatite slab clouds} 
\label{fig:spectral_fits_silicate}
\end{figure}

For Forsterite we retrieved an optical depth (tau at 1$\mu$m,) of 5.21$^{+0.90}_{-1.83}$. The optical depth retrieved (at 1$\mu$m) for the Enstatite cloud was surprisingly low with a logarithmic value of -2.36$^{+0.13}_{-0.13}$. This is much lower than the optical depth values derived in the \citet{Burningham_2021} and \citet{2023ApJ_Vos} studies -- see table \ref{tab:cloud_results_comparison}. A comparison of retrieved particle size distributions are indicated in Figure \ref{fig:cloud_property_illusration}. Despite the low optical depth  of enstatite, along with the forsterite, the cloud provides sufficient imprinting on the spectral shape to successfully fit the silicate feature in the MIRI data, as shown in Figure \ref{fig:spectral_fits_silicate}. 
For our iron deck we retrieved properties similar to that from \citet{Burningham_2021} for 2M2224 with the top of the deck positioned deep in the photosphere, as illustrated in Figure \ref{fig:bestfitTP}.

Figure \ref{fig:cloud_property_illusration} shows the particle size distributions for the clouds retrieved for VHS 1256 b. The values retrieved are akin to those seen in the \citet{Burningham_2021} and \citet{2023ApJ_Vos} study, where both Forsterite and Iron clouds have very small ($\leq$2$\mu$m) particles. The convergence towards such a small particle size can account for the small enstatite optical depth at 1$\mu$m. This preference for extremely small particle sizes, particularly the shape of our distributions as shown in Figure \ref{fig:cloud_property_illusration}, may be indicative that the Hansen profile parameterisation isn't capturing the true particle size distributions of cloud species present in the atmosphere of this object.

\begin{figure}
\centering
\includegraphics[width=0.37\textwidth]
{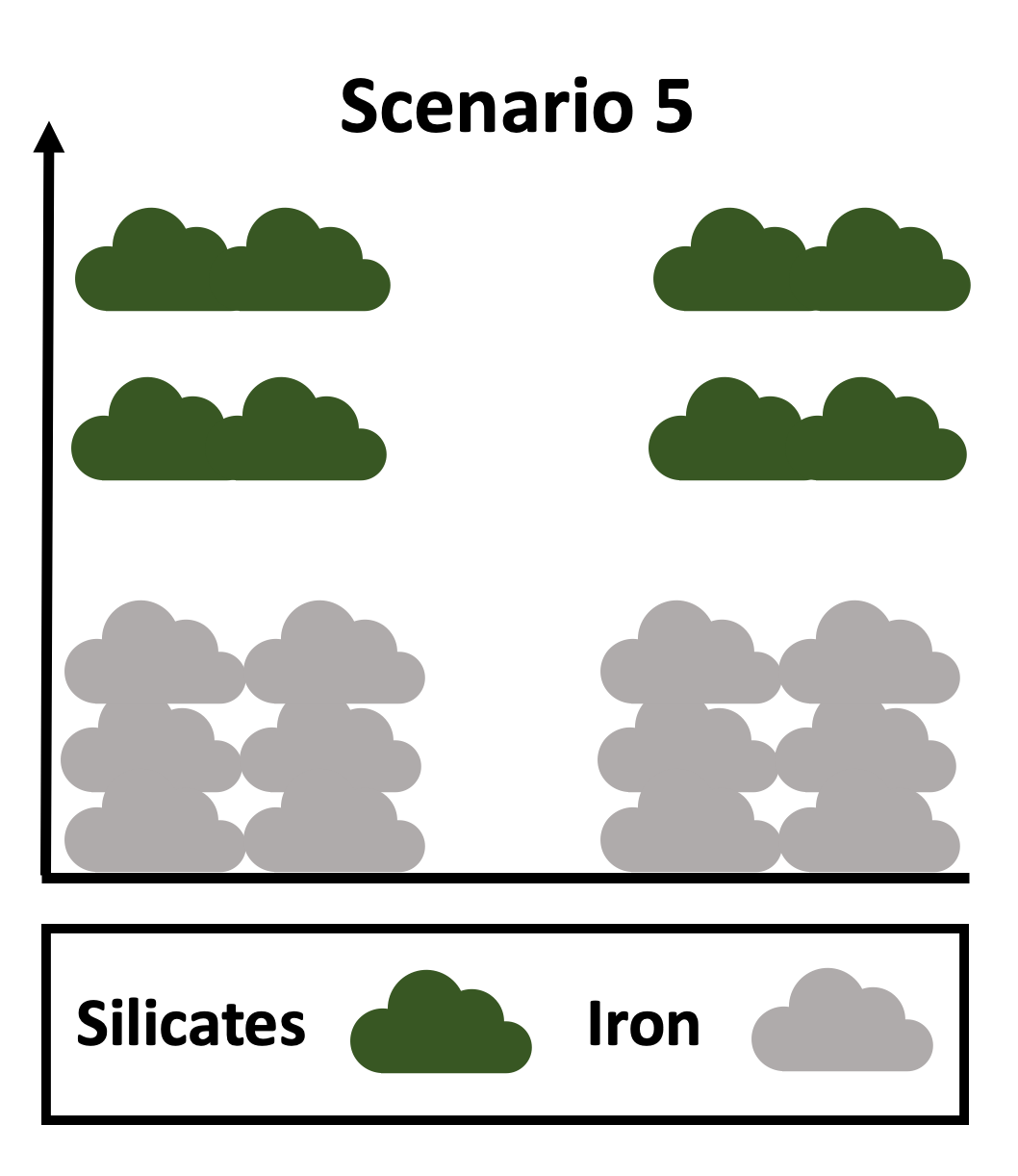}
\caption{This figure outlines a potential cloud scenarios we didn't investigate for VHS 1256 b.}
\label{fig:cloud_scenario_5}
\end{figure}

\subsection{Spectral fit}

The spectral fit of our retrieval is shown in Figure \ref{fig:spectral_fits}. As shown, generally, we achieve a successful fit (within $1\sigma$) across the 1-18 $\mu$m spectrum of VHS 1256 b. However, there are some wavelength regions that are poorly fit (outside $1\sigma$). The worst fitting region is in the L band  from $\sim$3.0$\mu$m--3.8um. We also see that the model retrieves a slightly offset continuum slope in the MIRI Channel 3 spectral window from $\sim$15$\mu$m--18$\mu$m. We acknowledge that mismatches may not only be model driven but also due to imperfect data reductions and systematics.

\subsection{Chemistry: H$_{2}$O, CO, CH$_{4}$, CO$_{2}$ and NH$_{3}$ constrained}

Within our retrieval analysis we included a list of 12 molecules for abundance measurements. Our best fit retrieval places tight constraints on H$_{2}$O, CO, CH$_{4}$, CO$_{2}$ and NH$_{3}$ -- see corner plots in Figure \ref{fig:chem_and_bulk_corner_plot}. The constraint on NH$_{3}$ is of interest for an L dwarf, with \citet{2022MNRAS_Suarez} showing NH$_{3}$ absorption onset in atmospheres with an early T classification. Thus, VHS 1256 b seems to be the  earliest in spectral sequence in which NH$_{3}$ has been seen or constrained by a retrieval. 

The rest of the molecules included in the retrieval (TiO, VO, CrH, FeH, H$_{2}$S Na, and K) are not tightly constrained. In the case of TiO and VO, which was successfully constrained in \citet{Burningham_2021}, it is likely that VHS 1256 b's observable atmosphere is too cool to host these species. CrH, FeH and Na+K are constrained in both \citet{Burningham_2021} and \citet{2023ApJ_Vos} but not here for VHS 1256 b. As in \citet{2023ApJ_Vos} we also don't see a tight constraint placed on SiO. H$_{2}$S also lacked a constraint in our retrieval. We note here, again, that we fit the data after convolving to resolution R of 300. This, combined with the presence of clouds, may act to damper or mute the signatures of the aforementioned molecules.

\begin{figure*}
\centering
\includegraphics[width=0.7\textwidth]
{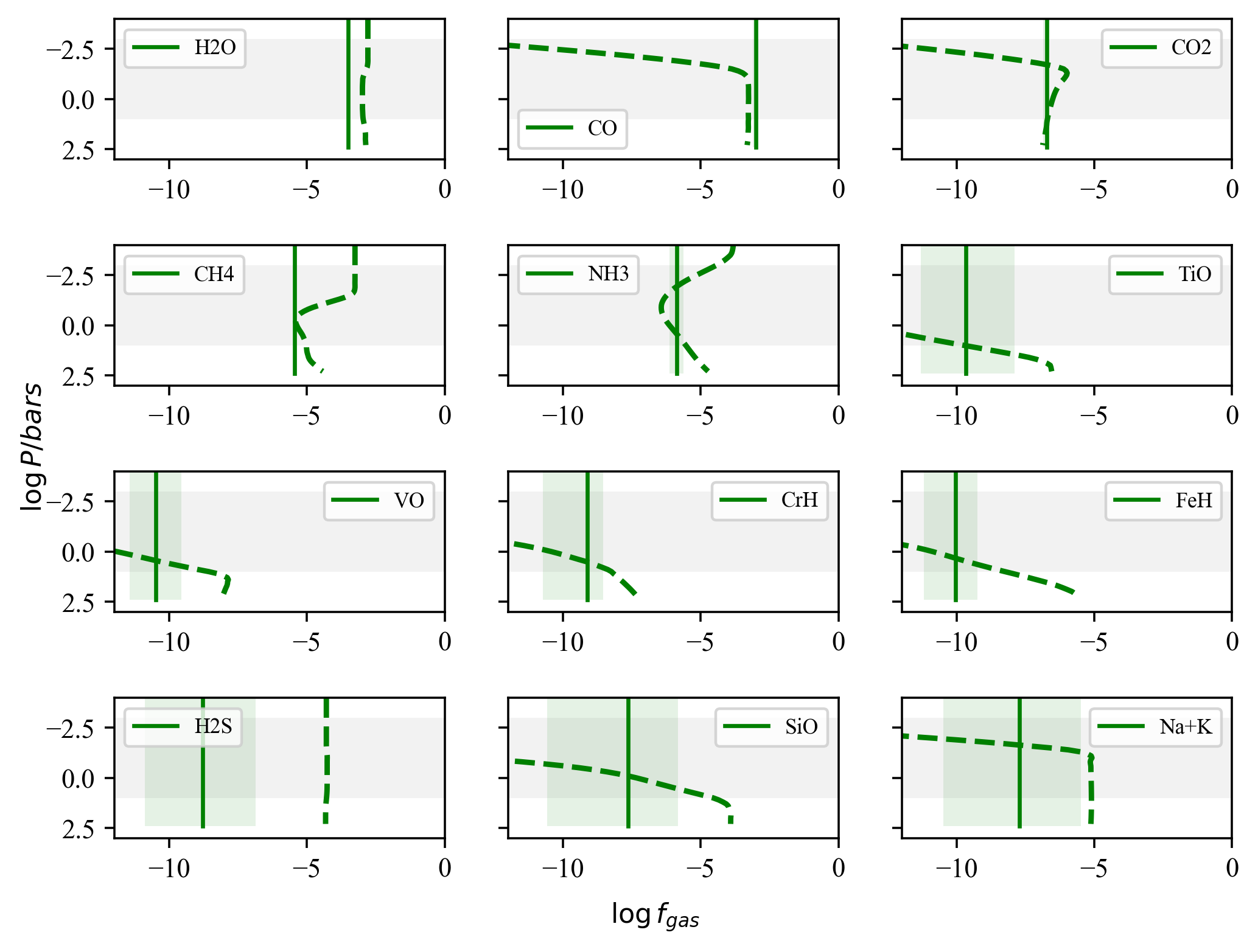}
\caption{Retrieved molecular vertical mixing ratios (green solid line) compared to the vertical mixing profiles predicted by a thermochemical equilibrium model (green dashed lines). Grey shading indicates the photospheric region.}
\label{fig:VMRs}
\end{figure*}

Figure \ref{fig:VMRs} shows the retrieved molecular abundances (which are constant in pressure space) with their uncertainties compared to the profiles predicted by a thermochemical equilibrium model. The rule of thumb is to expect the vertically constant mixing ratios to be representative of the mean photospheric abundance which is driving the retrieval fit. Thus, in the case of perfect equilibrium, our retrieved abundances should overlap with the profile from the thermochemical equilibrium model in the approximate region of the photosphere as indicated in Figure \ref{fig:VMRs}. For VHS 1256 b we expect a degree of disequilibrium chemistry to be present due to convective vertical mixing between the hotter (lower) CO and cooler (higher) CH$_{4}$ reservoirs, as was already suggested by the preliminary modelling performed in \citet{2023ApJ_Miles}. Figure~\ref{fig:VMRs} shows our retrieved CO abundance is slightly higher than the equilibrium model prediction while our CH$_{4}$ abundance is slightly lower than the equilibrium model prediction. We note that our assumption of our vertically constant mixing ratios restrict the models' ability to capture changes of abundances relative to pressure (altitude). CO$_{2}$ can also be used to trace disequilibrium chemistry \citep{2011_Burningham_ross458c}; however, our retrieved abundance of CO$_{2}$ aligns perfectly with the mean value of the equilibrium model. Overall, further investigation into, and testing of, these molecular abundances is required in order to determine if we are tracing indications of disequilibrium chemistry. We see that our retrieved NH$_3$ abundance matches that from the the equilibrium model. From the equilibrium model we also expect a reasonably high and detectable amount of H$_{2}$S and Na+K but neither of these abundances are constrained in our retrieval. For the rest of the trace species (TiO, VO, CrH, FeH and SiO) the lack of detections is consistent with the low abundances predicted by equilibrium models.

\subsection{Temperature-pressure profile}
\label{results:TP}

As outlined in Section \ref{sec:retrieval_recipe}, we used the \citet{jr:Mad&Seager2009} temperature-pressure profile parameterisation. We show the best fit retrieved profile in Figure \ref{fig:bestfitTP}. This profile is much less adiabatic and more isothermal in slope than predicted by self-consistent models such as the Sonora Diamondback grid \citep{2024_Morley}. This is particularly apparent in the photosphere and the deepest layers of the model. A non-adiabatic temperature-pressure gradient has regularly been retrieved for red L dwarfs \citep[e.g.,][]{Molliere_2020, 2024arXiv_Phillips} and is the atmospheric characteristic which cornerstoned the development of the ATMO diabatic convection (thermochemical instability) models \citep{2015_Tremblin}. 

Unlike in \citet{Burningham_2021} and \citet{2023ApJ_Vos}, the temperature-pressure profile we present for VHS 1256 b did not derive a gradient akin to that from the cloudy self-consistent models which was the initial expectation due to the similarly wide wavelength coverage. We do note that throughout the testing of retrievals on the VHS 1256 b data, we saw significant scatter and a lack of stability in the temperature-pressure profile we retrieved with examples of both more isothermal and more adiabatic profiles being seen. This has been a commonality of retrieval papers particularly when investigating the atmospheres of young low surface gravity objects \citep[e.g.,][]{2023MNRAS_Whiteford,2023AJ_Zhang,2024arXiv_Phillips}.

\subsection{Bulk properties}
For surface gravity we retrieve $\log g$ of 4.68$^{+0.03}_{-0.12}$ dex with retrieved radius 1.29$^{+0.02}_{-0.03}$Rjup and thus inferred a mass of 32.09$^{+2.16}_{-8.22}$Mjup. We note, however, that the values for mass, and thus $\log g$, are converging to the upper bound of the $\leq$35 Mjup parameter space of our retrieval. These mass and $\log g$ values are inconsistent with either of the the dual solutions outlined in \citealt{2023MNRAS_Dupuy} using the \citet{2008_Saumon_Marley_Evolution_models} models. 
Given the typical behavior of 1D retrievals to derive small radii ($\ll$1R$_{jup}$, see \citealt{Burningham_2021}, \citealt{2023ApJ_Vos}) it is surprising that we see such a sensibly high radius of 1.29$^{+0.02}_{-0.03}$Rjup for a young object such as VHS 1256 b. This, perhaps, may be a result of retrieving across such high quality observations and having data all the way out to 18$\mu$m. However, we acknowledge the true radius may be even higher due to VHS 1256 b's youth and low surface gravity, with such a value predicted by Exo-REM in \citet{2024ApJ_Petrus_2024} (see Table \ref{tab:bulk_results_comparison}). We note that our tightly constrained radius value is mostly 1-2$\sigma$ consistent with the diverse array of self-consistent modeled inferred values from \citet{2024ApJ_Petrus_2024} which are all outlined in (See Table \ref{tab:bulk_results_comparison}). 

Our retrieved approximately solar C/O (determined from the retrieved molecular abundances and correcting for oxygen depletion due to clouds as outlined in \cite{2024ApJ_Calamari}) for VHS 1256 b is also consistent at 2$\sigma$ with those from \citet{2024ApJ_Petrus_2024}, however, our [M/H] values are inconsistent. VHS 1256 b orbits a pair of M dwarfs \citep{2016_Stone} which can yield comparative bulk chemistry values -- and thus potential inference into formation history.  However we leave that exercise of host star abundance constraint and comparison for a future endeavor.

\section{RESULTS II: TESTING A COMPLEX DATASET}
\label{sec:results_II}

\begin{figure*}
\centering
\includegraphics[width=0.99\textwidth]{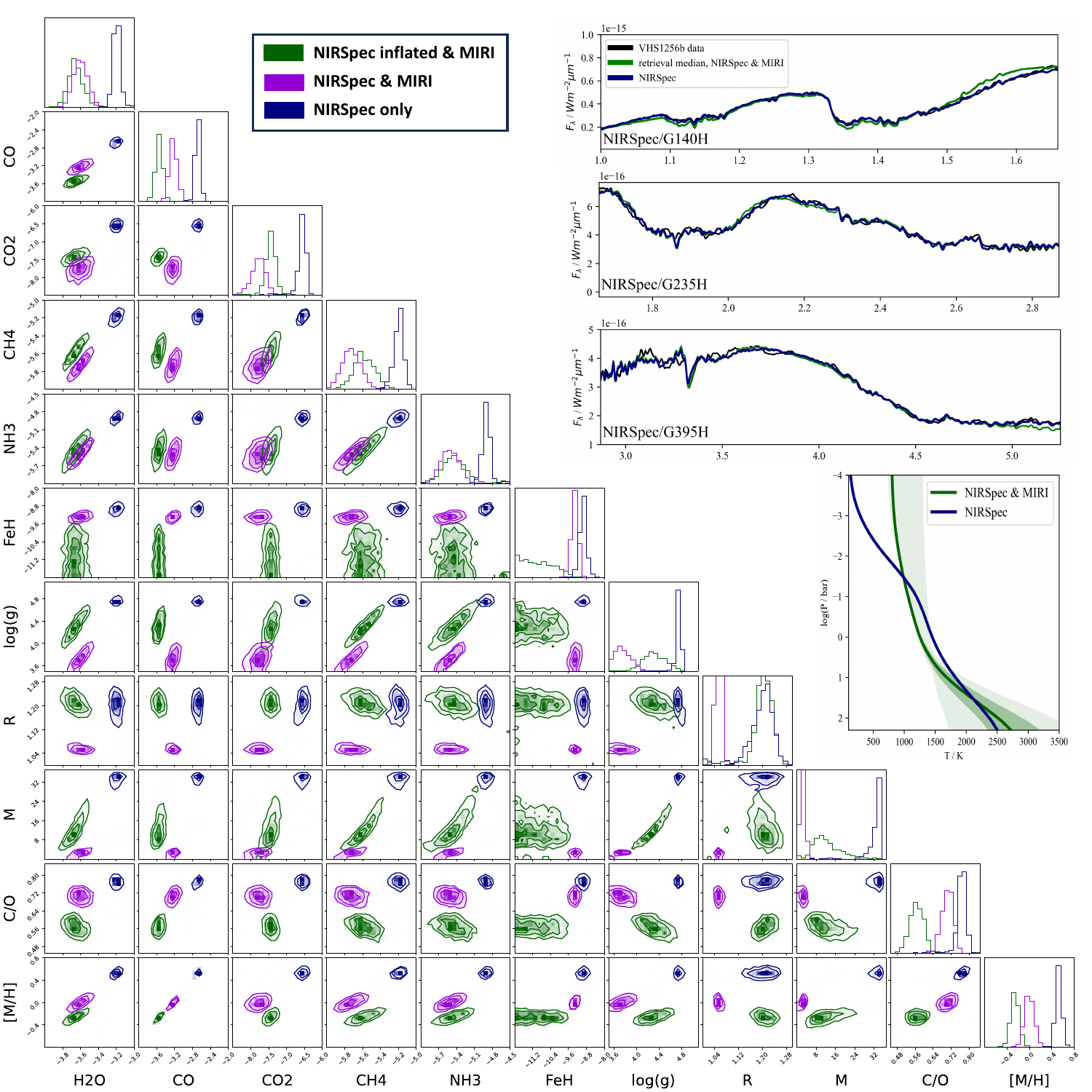}
\caption{Bottom left: Comparison corner plot of retrieved modelled parameters which are well constrained using 3 different data inputs. Top Right: Comparison of the retrieval spectral fit to the NIRSpec data when employing both NIRSpec and MIRI data vs only NIRSpec data. Middle right: Comparison of the retrieval temperature-pressure profile when when employing both NIRSpec and MIRI data vs. only NIRSpec data.}
\label{fig:posterior_comparison}
\end{figure*}

Here we outline a demonstration of what happens if we used the native SNR levels of the NIRSpec vs. MIRI data and, also, what happens if we used only NIRSpec data in our retrieval. This test was motivated by the novelty of the VHS 1256 b dataset and the understanding that retrieval results can produce precise but not necessarily consistent results when different data are used with the same model \citep{2023MNRAS_Whiteford} or the same data with differing models are employed \citep{2024ApJ_Nixon}. For this experiment we used Brewster as outlined in Section \ref{sec:retrieval_recipe} with a uniform forsterite and iron cloud model in order to compare the results of 3 different dataset scenarios. The data used were (1) NIRSpec and MIRI with NIRSpec's error bars inflated by an order of magnitude --  which was the data used for results outlined in Section \ref{sec:results} --, (2) NIRSpec and MIRI with no error bar inflation and (3) only the NIRSpec data. We neglected to consider only MIRI as 1-5$\mu$m is essential for chemical and temperature information thus necessitating near-IR data to be considered. Figure \ref{fig:posterior_comparison} illustrates the main findings from this test. Generally, we see a lack of overlap in the retrieved parameter posteriors, differing TP gradients and differing degrees of goodness-of-fit when we alter the SNR of the NIRSpec data or use only the NIRSpec data. A great deal of further testing is required to fully understand this behavior but these results suggests caution be taken not to over-interpret retrieval constraints in this early phase of model fitting to rich observations of objects such as VHS 1256 b.

\section{Outlook to the future: A new generation of retrieval studies for a new era of observations}

The exploratory and preliminary retrieval analysis conducted within this study is a small component of a larger modelling effort to holistically understand the multi-parameter nature of both brown dwarfs and directly imaged exoplanets. Motivated by the lessons learned from this work, such as the scatter of retrieval constraints demonstrated in Figure \ref{fig:posterior_comparison}, the following subsections discuss the next phase(s) of modelling for this dataset and datasets like this which will require a community-wide effort in the pursuit of precise, robust and accurate retrieval inferences.

\subsection{Full resolution, 1-20$\mu$m retrievals}

The next step will apply our retrieval framework to the data at their native resolution. We remind the reader that we convolved the NIRSpec and MIRI data from their native resolutions of R$\sim$3000 to R$\sim$300 for this study. This was mainly in an effort to reduce computational expense as an increase in resolution would greatly increase the retrieval run times. However, the possibility of using the full resolution data is a vital avenue that will be explored in the near future. 

\subsection{Samplers: Which to use?}

There are multiple sampler approaches currently being used in Bayesian retrieval frameworks for directly imaged exoplanets and brown dwarfs. The most commonly used sampler at the moment is MultiNest \citep{Feroz_Hobson_2008, Feroz_2009, PyMultiNest} with EMCEE \citep{2013_Foreman-Mackey} also being used in several studies, including this work. The decision to use EMCEE over MultiNest was motivated by the experiences of the \citet{Burningham_2021} study, where MultiNest struggled to converge/finish with such a highly dimensional (36-42) retrieval framework. This high dimensionality, compared to other retrieval works, is mainly due to complexity and flexibility of the cloud parameterisations available within the $Brewster$ framework. However, future work which tests our top few ranked models using different sampling algorithms, in order to check their respective retrieved values, will be a vital robustness check on retrieval recipes moving forward.
Other samplers such as PolyChord \citep{2015MNRAS_Handley_polychord}, which improves in efficiency as dimensionality is increased, may prove applicable to these kinds of retrieval comparison studies. An investigation, akin to the early retrieval methods testing of \citet{jr:LineRetrieval2013}, could explore the reproducibility of results when employing novel machine learning algorithms as well as classical Optimal Estimation approaches.

\subsection{Models Ranking: Which criterion to use?}

Despite the common use of BIC to compare and rank models recent works have called into question its applicability and suitability for investigating exoplanet atmospheres \citep[e.g.,][]{2023AJ_Welbanks,2025arXiv251000169T_Thorngren}. For example, \citet{2025arXiv251000169T_Thorngren} outlines the several drawbacks of the BIC metric and recommend such a criterion be used in parallel with the Akaike Information Criterion \citep[AIC,][]{1974ITAC_Akaike} or Bayesian Predictive Information Criterion Simplified \citep[BPICS,][]{Ando_2011}. Future work, using data such as that analyzed in this study, should explore employing these different metrics when conducting model comparisons.

\subsection{Demonstrating inference gains, loses and biases when combining JWST instrument modes}

We carried out an initial investigation with NIRSpec vs NIRSpec and MIRI in Section \ref{sec:results_II} but there is scope to dive much deeper. From the SED of VHS 1256 b it is clear that for objects in the cloudy L dwarf cohort that the bulk of the molecular signatures are encoded in the near-IR ($\leq5\mu$m) data while the silicate cloud features are encoded in the mid-IR ($\geq5\mu$m). Therefore, the results outlined in Section \ref{sec:results_II} show the importance of combining NIRSpec and MIRI data when constraining the global building blocks of such an atmospheres. 
The VHS 1256 b data from \citet{2023ApJ_Miles} will allow us to investigate inference modelling gains and losses across JWST's observing modes. Such a study would be able to demonstrate which data points drive which retrieved parameters \citep{2023AJ_Welbanks} and also demonstrate biases that may be introduced due to some observational orders/modes being missing. 

\subsection{A multi retrieval framework study}

In this study, we used the $Brewster$ retrieval framework, using a molecular recipe based on the studies of \citet{Burningham_2021} and \citet{2023ApJ_Vos}. 
There are, of course, a variety of modelling approaches within multiple retrieval codes and recipes \citep[e.g.,][]{Irwin_2008,jr:LineRetrieval2013,Harrington_BART_2016,2015_Benneke, 2017_MacDonald_POSEIDON, 2018_Gandhi_Hydra, 2019_Molliere_petitRADTRANS, 2020PLATONII_Zhang, Kitzmann_2020, 2022ApJ_Howe_apollo, 2022_Refaie_Taurex3p1, 2023AJ_Zhang}. The specific topic which warrants in-depth investigation is how differing retrieval recipes can lead to biases in the global results and conclusions. There are several key retrieval forward model corner stones, such as temperature-pressure parameterisations, that differ across codes. Different parameterisations may have subtle, but demonstrable, ways of biasing the parameters constrained within a retrieval. 

\subsubsection{Testing different temperature-pressure parameterisations}

For substellar atmospheres there has been a wide range of temperature-pressure parameterisations explored in the literature, with varying degrees of flexibility. This flexibility spans retrieving a combination of temperature-pressure points which are linked via some form of interpolation \citep[E.g.][]{Line_2015, jr:TauRex1, Kitzmann_2020}, multiple atmospheric zones with varying and more physically motivated temperature gradients \citep[E.g.][]{jr:Mad&Seager2009, Molliere_2020, 2023AJ_Zhang} or enforcement of radiative equilibrium \citep[E.g.][]{Lavie_2017_HELIOS}. The selection of which parameterisation is used within a retrieval framework can bias the overall results \citep{2023MNRAS_Whiteford}. 
The flexibility of the profile when analysing L type and L/T transition objects, such as VHS 1256 b, is also crucial due to a tendency to adopt a more isothermal profile, departing from the adiabatic profiles predicted by self-consistent modelling. In our study, we also see such behaviour. Therefore, a much more exhaustive study of the impacts of employing different temperature-pressure profile parameterisations with varying degrees of flexibility, warrants further investigation, particularly mapping its impact on retrieved chemistry and cloud properties.   

\subsection{Wider investigation of cloud species and structure}

We stress that the winning cloud model from this study is by no means the final word on the cloud species and structures present in the atmosphere of VHS 1256 b. In the early stages of analysis, when we binned the errors along with the resolution and didn't inflate the errors in NIRSpec relative to MIRI, we struggled to fit the silicate feature at all. Once we adapted the relative SNR, and achieved successful fits of this silicate feature in the MIRI data, we elected to investigate a relatively short list of cloud species and structures when compared to \citet{Burningham_2021}. In the longer term, a dataset such as that used here, warrants a much more extensive exploration more akin to the that from \citet{Burningham_2021}. Specifically, as outlined in e.g., \citet{2021_Luna}, there is an abundance of different condensate species to explore and search for in the wavelength range covered by MIRI. We can also explore more cloud structures, For example, there are other scenarios such as that illustrated in Figure \ref{fig:cloud_scenario_5} where we do not treat the iron deck as uniform.

A clear and crucial next step will be the expansion and development of the cloud structures used in retrieval codes and forward modelling frameworks. The tried and tested slab and deck prescriptions used in this study, while rather successful, are a 1D recipe for a dynamic 4D world (a dynamic 3D world which spectrally evolves in time). We refer the reader to \citet[][Figure 10]{2021MNRAS_Tan_showman}, which demonstrates how a 3D model may distribute clouds for a variable object like VHS 1256 b. JWST data will allow for wider application and development of multi-dimensional modelling retrieval frameworks which have previously been explored with $HST$ and $Spitzer$ observations of exoplanets \citep[e.g.,][]{2020MNRAS_Irwin}.  This, in turn, could help better capture the cloud structures present in the atmosphere of VHS 1256 b.  

\subsection{Contextualising this object}


As a component of the ERS program ($\#$1386), the full SED of VHS 1256 b was insightful into the capabilities of JWST on directly imaged exoplanets and brown dwarfs. This spectrum will be the subject of extensive analysis in the years to come. But investigating one object will not unlock the mysteries of the L/T transition, particularly for young and low surface gravity objects. Therefore, a crucial characterisation avenue will be the comparison and contextualisation to complimentary observations in the future, both of objects transitioning from L into T types and objects which bookend both sides of this transitional population of objects. 



\subsection{Testing near-IR reddening: clouds vs chemistry}

While clouds have long been assumed to control the significant near-IR colour of low surface gravity companions, another theory has been presented. In the second scenario, the convective instability triggered by the temperature-induced transition of photospheric CO to CH$_{4}$ could drive a reduced temperature gradient \citep{2015_Tremblin,Tremblin_2016, Tremblin_2017, Tremblin_2019} that acts to shape the near-IR SED in a similar way as the presence of clouds. Therefore, the JWST observations of VHS 1256 b offers an ideal test bed to explore this model degeneracy. Settling this debate will likely be aided by the abundance of time series data being obtained by JWST, particularly in the case of JWST variability monitoring of VHS 1256 b which includes time series NIRSpec/G395H data which spans both CO and CH$_{4}$ features simultaneously. 
Perhaps these observations will offer an insight into the transitional behavior of CO and CH$_{4}$ and how this does or does not correlate to the documented near-IR variability of VHS 1256 b.   

\subsection{A computational bottleneck: the dawn of the machine learning era?}

The previous subsections have outlined an extensive set of exploratory and investigative studies that will likely be conducted in the future. One of the biggest challenges is the computational intensity and expense of such analysis. 
This, therefore, begs the question: How do we tackle this computational bottleneck in the coming years? This will undoubtedly be a huge area of exploration and may propel machine learning approaches such as JAX \citep{jax2018github} into the forefront of the analysis of brown dwarf and exoplanet observations from JWST.    

\subsection{Probing the 4th dimension}


In the same way that contextualising the SED of VHS 1256 b with other objects belonging to the same extended family will be vital to better understanding its atmosphere, so too will be the unlocking of its 4D behaviour via time series observations. The retrieval work here is a precursor to the more complicated task ahead of probing and successfully modelling significant spectral variability known to be present in this object \citep{2020_Bowler_vhs1256b, 2020_Zhou_vhs1256b, 2022AJ_Zhou}. JWST's early cycles will see a large amount of such time series datasets including for VHS 1256 b ($\#$3375, PI Whiteford).
These will, undoubtedly, revolutionise the classic 1D retrieval recipes that we currently employ and use in this study.

\section{SUMMARY AND CONCLUSIONS}

The overall aim of this work is to apply and test a relevant and complex retrieval framework to the spectrum of VHS 1256 b. Specifically the aim was  to probe the chemical and cloudy nature of this object across the full 1-18$\mu$m ERS \#1386 dataset. We started by outlining the challenges that are presented by the JWST observations of VHS 1256 b and other such objects observed with the same strategy in the early stages of the JWST mission. These include: 

\begin{itemize}
  \item The JWST ERS spectrum of VHS 1256 b is the first data of its kind to combine overlapping spectral orders. This introduces the issue of imperfect absolute calibrations,  biases, and the ability of overlapping regions (if included) to unevenly weight the retrieval. 
  \item Modelling across 1-18$\mu$m at the native JWST NIRSpec and MIRI MRS resolution orders presents a computational issue, adding to the already long retrieval runtime.  
  \item VHS 1256 b is one of the most dynamic objects of its kind. It presents the highest degree of spectral variability seen across imaged exoplanets and brown dwarfs. Its temperature places it in the camp of objects we predict to have a significant degree of vertical mixing. As such, we expect the atmosphere to have 4D dynamic properties and signatures while our retrieval framework only employs a traditional 1D approach.   
\end{itemize}

In this work we outlined how we tackled and navigated these challenges with our $Brewster$ retrieval framework setup. In order to make this initial retrieval effort computationally feasible, we convolved the native JWST data to a resolution of R=300 and used only one order in the zones of overlap. To avoid a higher weighting of the NIRSpec data, we deflated the SNR of the NIRSpec data (inflating the error bars by an order of magnitude) to more evenly match that of MIRI.  

We explored a list of cloud model combinations based on the results of \citet{Burningham_2021} and \citet{2023ApJ_Vos}. This included uniform and patchy combinations of MgSiO$_{3}$, Mg2SiO$_{4}$, SiO and Fe. In summary, The results of this analysis were: 

\begin{itemize}
  \item A combination of forsterite and enstatite were the top ranked cloud species for fitting VHS 1256 b's silicate feature. 
  \item Patchy clouds where significantly preferred over uniform clouds with a $\Delta$BIC of 297.  
  \item The retrieval places constraints on the presence of H$_{2}$O, CH$_{4}$, CO, CO$_{2}$ and NH$_{3}$ resulting in a C/O ratio slightly above solar and a [M/H] twice that of solar.
  \item We retrieved a high radius value of 1.29$^{+0.02}_{-0.03}$Rjup, agreeing with self-consistent forward models, with  while log(g) and mass converged to the upper end of the parameter space. 
  \item Our retrieved temperature-pressure profile was more isothermal (diabatic) in shape than the self-consistent modelling predictions.
  \item We achieved a successful fit, generally within 1$\sigma$, to the JWST spectrum of VHS 1256 b. There is some spectral structure that may not be successfully fit due to it being the result of systematic imprints of the data reduction pipeline.
\end{itemize}

In an effort to test our approach of degrading the NIRSpec SNR we also tested the scenarios of using NIRSpec plus MIRI data with raw SNRs as well as NIRSpec data with its raw SNR without MIRI data. This showed that the retrieved parameters are sensitive to the data scenarios explored. These results point to the need for caution when trusting the accuracy of highly precise retrieval parameter inferences. 

As such, with the experiences of all the retrievals performed for this study, we subsequently outline an extensive list of next steps with VHS 1256 b and its JWST observations. This includes exploring a full (native JWST) resolution retrieval, the need to test different retrieval frameworks against each other (particularly regarding temperature-pressure profile and chemistry assumptions), carrying out a more extensive clouds species and structure comparison (akin to \citealt{Burningham_2021}), tackling the computational bottleneck of retrievals with machine learning and, finally, investigating the nature of this dynamic 4D world using time-series observations.

\section{Acknowledgments}
This project was supported by a grant from STScI (JWST ERS-01386) under NASA contract NAS5-03127. This work benefited from the 2022 Exoplanet Summer Program in the Other Worlds Laboratory (OWL) at the University of California, Santa Cruz, a program funded by the Heising-Simons Foundation. JF acknowledges funding from the Heising Simons Foundation as well as NSF award \#2238468, \#1909776, and NASA Award \#80NSSC22K0142.
BB acknowledges support from UK Research and Innovation Science and Technology Facilities Council [ST/X001091/1].

\appendix

\section{Appendix information}

\begin{figure*}
\centering
\includegraphics[width=0.99\textwidth]
{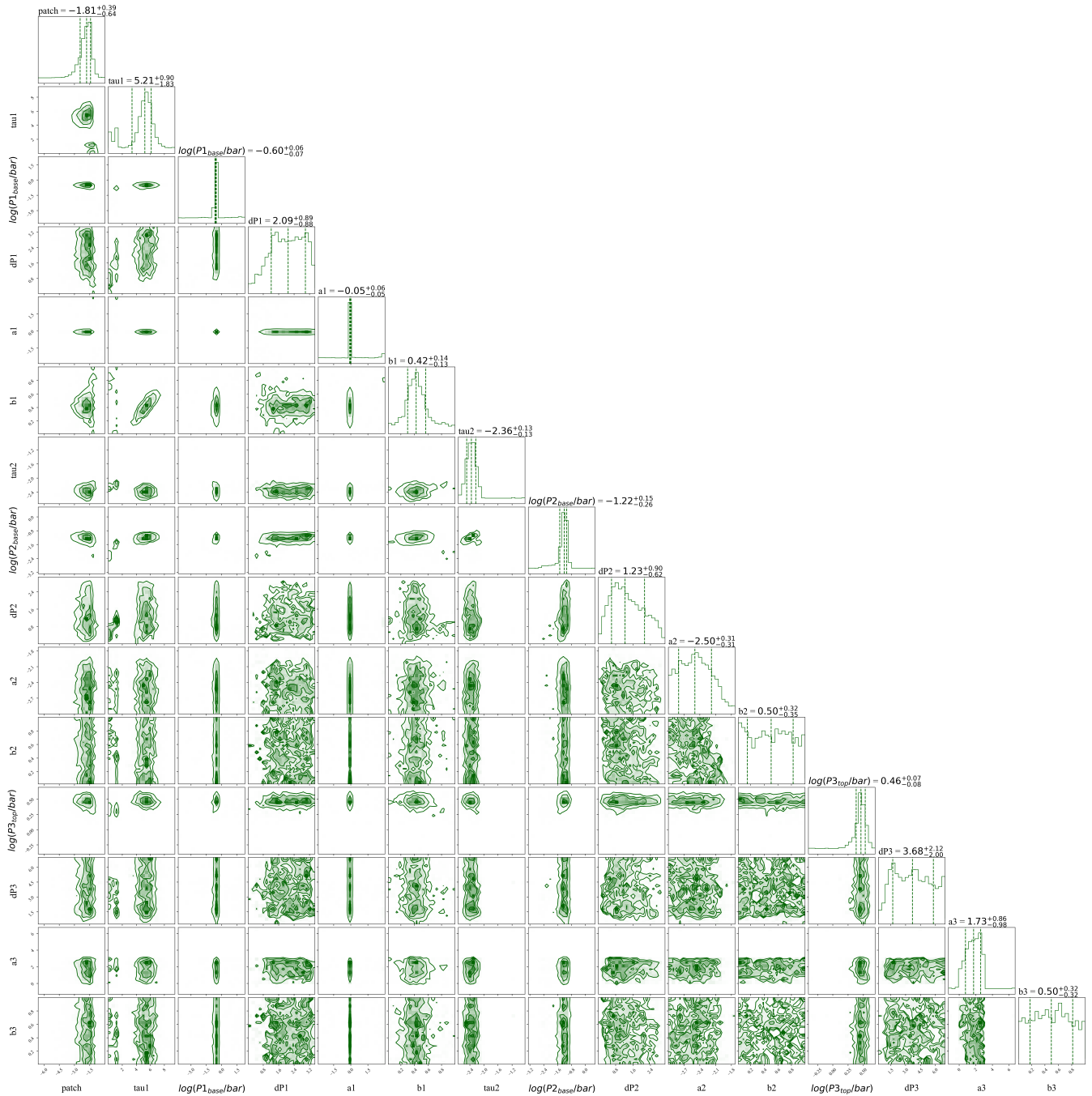}
\caption{Corner plot of retrieved cloud parameter values for the preferred model fit to VHS 1256 b. (1) denotes the parameters of the Mg$_2$SiO$_4$ slab, (2) denotes the parameters of the MgSiO$_43$ slab and (3) denotes the parameters of the Fe deck. Also shown is the cloud-free patch fraction in log space.}
\label{fig:cloud_corner_plot}
\end{figure*}

\bibliography{sample631.bib}{}
\bibliographystyle{aasjournal.bst}



\end{document}